\documentclass{aastex631}

\usepackage{prettyref}

\newrefformat{cap}{\hyperref[#1]{Figure~\ref{#1}}}
\newrefformat{fig}{\hyperref[#1]{Figure~\ref{#1}}}
\newrefformat{suppfig}{\hyperref[#1]{Supplementary Figure~\ref{#1}}}
\newrefformat{tab}{\hyperref[#1]{Table~\ref{#1}}}
\newrefformat{sec}{\hyperref[#1]{Section~\ref{#1}}}
\newrefformat{sub}{\hyperref[#1]{Section~\ref{#1}}}
\newrefformat{subsec}{\hyperref[#1]{Section~\ref{#1}}}
\newrefformat{cha}{\hyperref[#1]{Chapter~\ref{#1}}}
\newrefformat{lis}{\hyperref[#1]{Listing~\ref{#1}}}
\newrefformat{eq}{\hyperref[#1]{Eq.~\ref{#1}}}

\usepackage{amsmath}
\DeclareMathOperator\erf{erf}

\usepackage{newfloat}
\DeclareFloatingEnvironment[name={\textbf{Supplementary Figure}},fileext=lof]{suppfigure}

\begin{document}

\title{Evaluating the statistical similarity of neural network activity and connectivity via eigenvector angles}

\correspondingauthor{Robin Gutzen}

\author[0000-0001-7373-5962]{Robin Gutzen}
\affiliation{Institute of Neuroscience and Medicine (INM-6) and Institute for Advanced Simulation (IAS-6) and JARA-Institute Brain Structure-Function Relationships (INM-10), \\ 
J{\"u}lich Research Centre, J{\"u}lich, Germany
}
\affiliation{Theoretical Systems Neurobiology, \\ 
RWTH Aachen University, Aachen, Germany}

\author[0000-0003-2829-2220]{Sonja Gr{\"u}n}
\affiliation{Institute of Neuroscience and Medicine (INM-6) and Institute for Advanced Simulation (IAS-6) and JARA-Institute Brain Structure-Function Relationships (INM-10), \\ 
J{\"u}lich Research Centre, J{\"u}lich, Germany
}
\affiliation{Theoretical Systems Neurobiology, \\ 
RWTH Aachen University, Aachen, Germany}

\author[0000-0003-1255-7300]{Michael Denker}
\affiliation{Institute of Neuroscience and Medicine (INM-6) and Institute for Advanced Simulation (IAS-6) and JARA-Institute Brain Structure-Function Relationships (INM-10), \\ 
J{\"u}lich Research Centre, J{\"u}lich, Germany
}



\begin{abstract}
Neural systems are networks, and strategic comparisons between multiple networks are a prevalent task in many research scenarios. In this study, we construct a statistical test for the comparison of matrices representing pairwise aspects of neural networks, in particular, the correlation between spiking activity and connectivity. The "eigenangle test" quantifies the similarity of two matrices by the angles between their ranked eigenvectors. We calibrate the behavior of the test for use with correlation matrices using stochastic models of correlated spiking activity and demonstrate how it compares to classical two-sample tests, such as the Kolmogorov-Smirnov distance, in the sense that it is able to evaluate also structural aspects of pairwise measures. Furthermore, the principle of the eigenangle test can be applied to compare the similarity of adjacency matrices of certain types of networks. Thus, the approach can be used to quantitatively explore the relationship between connectivity and activity with the same metric. By applying the eigenangle test to the comparison of connectivity matrices and correlation matrices of a random balanced network model before and after a specific synaptic rewiring intervention, we gauge the influence of connectivity features on the correlated activity. Potential applications of the eigenangle test include simulation experiments, model validation, and data analysis.

\end{abstract}



\section{Introduction}\label{sec:intro}
Many processes in our world are networks \cite[see, e.g.,][]{Albert2002_47} consisting of connected, interacting nodes that exhibit a joint dynamics determining the network activity. This holds especially for the field of neuroscience, where networks describe structural architecture and information processing in the brain. A question that may arise in the analysis of such networks is to quantify the degree of similarity of them. Such comparisons of similarity could be limited to the connections between nodes, ignoring the network activity. Alternatively, one may ask if two networks are similar with respect to measured features describing their activity. Lastly, combining the former two approaches, one can investigate the influence of a change in connectivity on the change in activity.

When investigating the structure of connectivity and coordinated activity of a neural system in experiment or simulation we may find ourselves in a situation where we would like to compare specific realizations of different networks representing that neural system. The need for such comparisons could arise in a variety of contexts, e.g., to differentiate networks exhibiting certain features of the connectivity, to measure differences in the observed activity correlations, or to better understand the link between connectivity and activity.

Practically, performing comparisons of networks requires the identification of characteristic measures from the network that capture the relevant signatures of connectivity and activity of neural systems. The characterization of networks is then often accompanied by a further statistical evaluation of their similarity or dissimilarity based on the chosen measures. The combination of these two steps represents the core of a validation process \citep{Schlesinger1979_103} for network models which provides a formalized approach to the quantitative evaluation of network differences. Common validation scenarios include measuring the predictive power of a model with respect to experimental data, quantifying the variability between two experimental datasets \citep{Mochizuki2016_5736}, benchmarking the influence of parameter variation on a model \citep{Dasbach2021_757790}, or testing the agreement between two models \citep{Trensch2018_81, Gutzen2018_90, vanAlbada2018_291}.

A wide range of applications of graph-theoretic measures has been proposed \citep{Bullmore2009_186} to describe structural and functional networks, i.e., networks where the links between nodes represent either physical connectedness or functional similarity on the basis of the dynamics. While applicability of these measures, in particular, for characterizing non-linear aspects of the dynamics, is highly dependent on the network at hand \citep{Curto2019_11}, there are promising approaches derived from graph theory, e.g., quantifying functional similarity \citep{Haber2020_2020.04.27.057752}, measuring a neuron's impact on its surrounding network \citep{Vlachos2012_e1002311}, or establishing relations between structural motifs and moments of the correlation \citep{Pernice2011_e1002059}.

In the following, we focus on a generic case that considers characteristic measures for networks of spiking neurons. To describe the statistical properties of spiking activity, the characterization can be either based on single-neuron measures (e.g., firing rate), pairwise measures (e.g., Pearson correlation), or higher-order measures (e.g., graph transitivity). The single-neuron measures are the most straight-forward to use, as they are statistically independent and can be compared by a range of suitable two-sample tests. Popular tests are, for example, the Student's t-test \citep{Student1908_1} or the Kolmogorov-Smirnov test \citep{Hodges1958_469}, and, even though they are not formally statistical tests, also comparative measures such as the Kullback-Leibler divergence \citep{Kullback1951_79} and the effect size \citep{Cohen1988_a}.

The single-neuron measures, which only capture a limited set of statistical features characterizing the network dynamics, can be complemented by pairwise (and higher-order) measures of the interdependence between neurons. Although distributions of such pairwise measures may also be compared using two-sample tests by reducing, for example, a correlation matrix to a distribution of correlation coefficients, the sample values in this case are not statistically independent and information is lost in the reduction from a 2D measure to 1D. Hence, comparing matrices representing the pairwise measure in a meaningful way is not trivial, and only few statistical tests are available \citep{Flury1988_, Krzanowski1990_81, Calsbeek2009_2627}. Approaches to perform comparisons between matrices that do not involve a statistical test include using the correlations between matrix elements, the Euclidean distance, or the geodesic distance \citep{Venkatesh2019_116398}.

We here construct a statistical test to compare pairwise measures of two networks, in particular the matrices of Pearson correlation coefficients, to evaluate network similarity in more detail. The eigenvectors of a matrix span the space in which the data in the matrix is represented most naturally, in the sense that the first eigenvector points along the direction of the largest variance in the data, the second along the direction of the largest variance within the orthogonal subspace, etc. The corresponding eigenvalues quantify the variance along these axes. In the case of a correlation matrix, this means that the first eigenvectors points towards the dominant feature of the correlation structure, which could be, for example, a strongly correlated group of neurons. Thus, to measure the similarity between two matrices we choose to evaluate the alignment of the respective eigenvectors. To this end, we evaluate the angles between them (which we term "eigenangles") and how they behave for the case of independent random matrices. In order to properly define such angles, we construct the test with the assumptions of large matrices that are defined in the same space, i.e., two correlation matrices that are calculated from the same set of neurons.

In the following, we will build up a theoretical basis for the statistical test by first describing the behavior of angles for high dimensional random vectors, and the special case of eigenvectors. Then, we will look at a measure we term "angle-smallness" as an indicator for their similarity and integrate the eigenvalues as a weighting factor into the analytical description. Based on this, we formulate a similarity score for the matrix comparison and describe its analytical distribution to compute a corresponding $p$-value. We then characterize the statistical test by applying it to calibration scenarios of stochastic and simulated neural network activity. We further explore the extension of the test to pairwise measures that capture the connectivity of the network, i.e., the adjacency matrix of synaptic weights. In this way, we demonstrate how this approach can be used to draw links between network connectivity and dynamics on the basis of a set of network simulations where the connectivity is modified.

\section{Statistical eigenangle test for correlation matrices} \label{sec:statistical-eigenangle-test}

Our objective is to quantify the similarity of a pair of matrices by means of a statistical test based on the angles between their eigenvectors. Throughout this section, we will assume the case of comparing two correlation matrices as illustrated by the concrete use case of correlations between $N$ neuronal spike trains. We generalize the concept to a more generic class of matrices, such as graph adjacency matrices, in the later part of this report.

Let $\{\mathbf{v}^A_i, i\in 1,\dots,N\}$ denote the set of ordered, normalized eigenvectors of a matrix $\mathbf{A}$, such that $\lambda^A_i\ge\lambda^A_{i+1}$ for each corresponding eigenvalue $\lambda^A_i$. We define the $i$-th \textit{eigenangle} $\phi_i$ as the $\phi_i= \measuredangle \ ( \mathbf{v}^A_i, \mathbf{v}^B_i) = \arccos(\mathbf{v}^A_i \cdot \mathbf{v}^B_i) \in [0,\pi]$ between the eigenvectors $\mathbf{v}^A$ and $\mathbf{v}^B$ of matrices $\mathbf{A}$ and $\mathbf{B}$, respectively. Thus, we consider the angles between pairs of the $i$-th ordered eigenmodes of each of the two matrices.

\begin{figure}[!h]
	\centering
	\includegraphics[width=.85\textwidth]{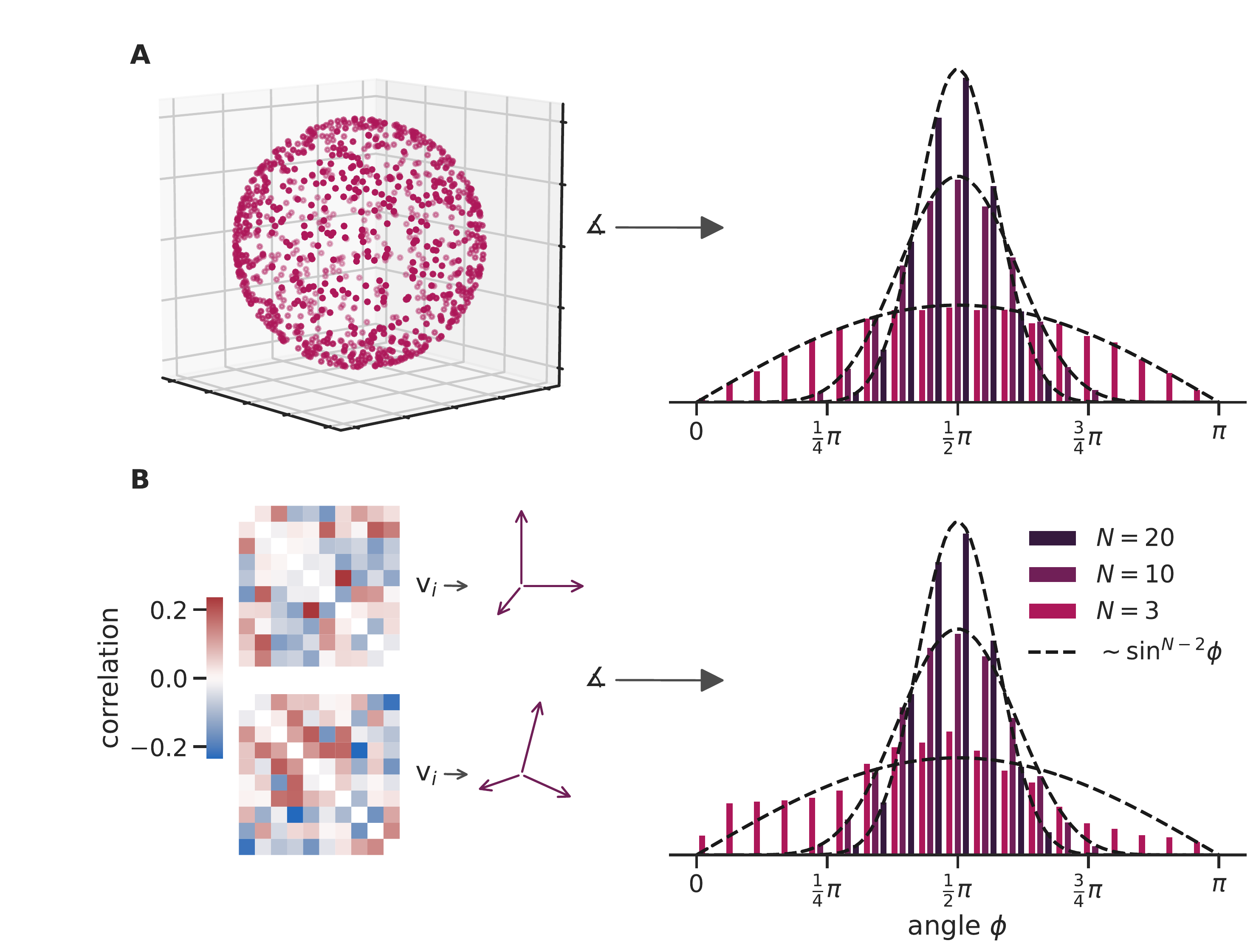}
	\caption{\textbf{Distributions of angles between vectors.} \textbf{A)} Examples of distributions of angles between normalized random vectors for dimensions $N=3$, $N=10$, and $N=20$ (colors of the histogram). Dashed curve: analytical distribution proportional to $\sin(\phi)^{N-2}$. \textit{Left:} example of random vectors on a sphere in $N=3$. \textbf{B)} Examples of distributions of angles between the corresponding eigenvectors $\mathbf{v}_i$ of two random correlation matrices. Dashed curve: analytical prediction for random vectors as in panel A. \textit{Left:} Example realization of two correlation matrices for $N=10$. Each histogram contains $10^4$ angles from the corresponding sampled vectors.}\label{fig:fig1}
\end{figure}

The underlying assumption for the eigenangle test is that a small angle between two eigenvectors indicates similarity of the corresponding eigenmode, whereas a near-orthogonal angle indicates a discrepancy. Following this assumption, we will derive criteria to identify similar eigenmodes based on the eigenangles $\phi_i$ by calculating their expected distribution under the assumption of independent activity. Let us consider a set of $N$-dimensional normalized random vectors, i.e., vectors that are uniformly distributed over the $N$-dimensional unit sphere. Here, by assuming matrices $\mathbf{A}$ and $\mathbf{B}$ to be correlation matrices containing the correlation coefficients between individual neurons in the network, the dimensionality $N$ of this space is equal to the number of neurons we record from.
Contrary to the intuition suggested by a unit circle in $N=2$, in higher dimensional spaces ($N>2$) the probability distribution $f_\measuredangle(\phi)$ of angles between two random vectors is not uniform. To illustrate this dependency, let us imagine a point at the pole of the unit sphere (i.e., the intersection of the sphere with the first axis) and a randomly chosen second point on the unit sphere. For $N=2$, each section of the circle $d\Theta$, with angle $\theta$ from the pole, is equally likely to contain the second point. Thus, the probability distribution of the angle between two arbitrary vectors will be uniform. However, for $N=3$ the surface of the ring section $d\Theta$ under the polar angle slice $d\theta$ increases with distance $\theta$ from the pole and is maximal at the equator. Therefore, the probability to observe an angle between two random vectors will be increased around $\pi/2$, i.e., the vectors tend to be perpendicular (cf., \prettyref{fig:fig1}A). The same effect is observed for larger dimensions $N>3$. The probability distribution can be approximated numerically by sampling the angles between a set of random vectors, created by drawing the vector components independently from a normal distribution and then normalizing the vectors \citep{Guhr1998_189}. Indeed, the probability distribution can also be calculated analytically \citep{Cai2013_1837}:
\begin{equation}
f_\measuredangle(\phi) = \frac{\Gamma(\frac{N}{2})}{\sqrt{\pi} \Gamma(\frac{N-1}{2})} \sin(\phi)^{N-2} \qquad \phi \in [0,\pi]
\label{eq:angle-distribution}
\end{equation}

In the context of this study, however, we consider angles between eigenvectors which are not randomly distributed, but instead are pairwise orthogonal due to correlation matrices being real and symmetric. We can numerically demonstrate that the effects of these additional constraints are only relevant for low-dimensional spaces and that the distribution of eigenangles can be approximated by $f_\measuredangle(\phi)$ for higher dimensions. To show this, we first define a random correlation matrix $\mathbf{A}$, which is positive definite, symmetric, whose elements are real-valued random variables $A_{j,i} \in$ [$-1,1$], and diagonal elements are $A_{i,i}=1$. A random correlation matrix can be created by calculating the Gram matrix from a set of normalized random vectors $\mathbf{Y} = \mathbf{X}\mathbf{X}^*$, where $\mathbf{X}$ is a matrix with rows $X_k$ being normalized random vectors and $\mathbf{X}^*$ denoting the conjugate transpose of $\mathbf{X}$ \citep{Holmes1991_239}. The dimension of the row vectors $X_k$ does not influence the distribution of the eigenvectors, and therefore, we describe this degree of freedom as $\alpha \cdot N$. While $\alpha$ has no relevance for the distribution of eigenangles, it will influence the distribution of eigenvalues described below. \prettyref{fig:fig1}B demonstrates a representative example of how the distribution of angles between eigenvectors are well approximated by the analytic distribution $f_\measuredangle(\phi)$ of angles between random vectors for higher dimensions (about $N>10$). Therefore, using $f_\measuredangle(\phi)$ as an analytical approximation is appropriate for our approach if we consider the analysis of correlations between large numbers of neurons (e.g., $N>100$, which describes a common scenario in the analysis of electrophysiological data).

Since we motivated the use of eigenangles to indicate similarity, we define the deviation from orthogonality towards small angles $\Delta_i = 1 - \frac{\phi_i}{\pi/2}$ as the auxiliary variable \textit{angle-smallness} to quantify the similarity of two vectors on a scale from $-1$ to $1$. Performing a variable transformation on the random angle distribution (\prettyref{eq:angle-distribution}), we obtain the corresponding distribution for the angle-smallness:
\begin{equation}
\tilde{f_\measuredangle}(\Delta) \propto \cos^{N-2}(\Delta\cdot\pi/2), \qquad \Delta \in [-1, 1]
\label{eq:smallness-distribution}
\end{equation}

The eigenvalues of positive definite matrices provide a measure to describe the amount of variance captured by the relative contributions of individual neurons to the corresponding eigenvectors. For that reason, we argue that eigenvectors with larger eigenvalues have a more dominant role in defining the structure of the correlation matrix as opposed to eigenvectors with small eigenvalues. Thus, in designing a cumulative test score to quantify the similarity of two correlation matrices taking into account all pairs of eigenvectors, those with the highest eigenvalues should be weighted stronger than those with low eigenvalues. To incorporate this aspect into the test, we weight the angle-smallness $\Delta_i$ between $i$-th eigenvectors of two matrices $\mathbf{A}$ and $\mathbf{B}$ with the quadratic mean of the corresponding eigenvalues 
$w_i \propto \sqrt{({\lambda^A_i}^2 + {\lambda^B_i}^2) / 2}$.

In a next step, we derive the analytic distribution of the weighted angle-smallness $w_i\Delta_i$ for independent random matrices. The Marcenko-Pastur distribution \citep{Marcenko1967_457} given by
\begin{equation}
h_{\alpha}(\lambda) = \frac{\alpha}{2 \pi \lambda} \sqrt{(\lambda_+ - \lambda) \cdot (\lambda - \lambda_-)}
\\
\lambda_{\pm} = \left(1 \pm \sqrt{\frac{1}{\alpha}}\right)^2
\label{eq:MP-distribution}
\end{equation}
describes the distribution of eigenvalues $\lambda$ for matrices of the type $\mathbf{Y}_{N}=\mathbf{XX}^{T}$, where $\mathbf{X}$ is an $(\alpha N)\times N$ random matrix whose entries are independent identically distributed random variables with mean $0$ and variance $\sigma^{2}<\infty$, and $\mathbf{X^T}$ its transpose. The distribution is asymptotic for $N\rightarrow\infty$, and can be considered a good estimate for the scenario of $N>100$ considered in this study. The distribution $h_{\alpha}(\lambda)$ only depends on the parameter $\alpha$ which we introduced above as the ratio between the length of the row vectors $X_k$ and the dimensionality $N$. In the concrete application of $\mathbf{Y}_N$ representing a matrix of correlation coefficients between spike trains, the row vectors $X_k$ correspond to the binned spike trains, so that the parameter $\alpha$ is determined by the number of bins divided by the number of spike trains.

By defining the weights $w_i$ as the quadratic mean of the eigenvalues $\lambda^A_i$ and $\lambda^B_i$ they also follow the Marcenko-Pastur distribution. Combining \prettyref{eq:smallness-distribution} and \prettyref{eq:MP-distribution}, we can formulate the distribution of the weighted angle-smallness $w\Delta$ as
\begin{equation}
g_{N,\alpha}(w\Delta) = \int_{\lambda_-}^{\lambda_+} \tilde{f_N}(\frac{w\Delta}{\lambda}) \cdot h_{\alpha}(\lambda) \cdot \frac{\mathrm{d}\lambda}{\lambda}
\label{eq:weighted-angle-smallness-distribution}
\end{equation}
under the assumption of two independent random correlation matrices. We note that in this formalism alternative choices for the weights $w$ are possible as long as there is a corresponding analytical description for their distribution. 

Given the $N$ individual values of the weighted angle-smallness obtained by comparing corresponding eigenvectors of the two matrices, we define the scalar \textit{similarity score}
\begin{equation}
\eta = \frac{1}{N} \sum_i^N w_i\Delta_i
\end{equation}
as their average. A large positive $\eta$ therefore indicates that the angles between the eigenvectors (in particular, those corresponding to the largest eigenvalues) tend to be smaller than expected for independent random matrices. This indicates that the two matrices have common non-random structures. In order to interpret a given value of $\eta$ for a sample with given $N$ and $\alpha$, we derive how $\eta$ is distributed under the null hypothesis of random matrices. According to the central limit theorem, for large $N$ the average of the weighted angle-smallness $w \Delta$ will converge towards a normal distribution centered around the expected mean (here $0$) with the standard deviation $\sigma=s/\sqrt{N}$, where $s$ is the standard deviation of $g_{N,\alpha}(w\Delta)$. Expressing $s$ as the integral over the product of the distribution with the squared distance to the mean, we obtain the final analytical description of the distribution of the similarity score
\begin{equation}
f(\eta) = \frac{1}{\sqrt{2\pi\sigma^2}} \exp(-\frac{\eta^2}{2\sigma^2}) \label{eq:similarity-distribution}
\end{equation}
where
\begin{equation*}
\sigma^2 =  \frac{1}{N} \int x^2 \cdot g_{N,\alpha}(x) \ \mathrm{d}x .
\end{equation*}

Based on \prettyref{eq:similarity-distribution}, we define the null distribution in the context of a Null Hypothesis Significance Test (NHST). As the construction of $f(\eta)$ was based on the assumption of independent random matrices, the null hypothesis could be best expressed as "the two matrices have no shared non-random structures". We assumed the matrices to be of the type $\mathbf{Y}_{N}=\mathbf{XX}^{T}$. Moreover, we assumed that $N$ is large to account for the constraint that eigenvectors are orthogonal in  \prettyref{eq:angle-distribution}, to be able to apply the central limit theorem, and since the integrated Marchenko-Pastur distribution is asymptotic for $N \rightarrow \infty$. From numerical simulations we found that $N>10$ is sufficient so that the null distribution $f(\eta)$ reasonably represents the randomly sampled test data.

Violations of the null hypothesis, in particular any substantial correlation between the matrices, result in a score so large that it is unlikely to be explained just by the width of the null distribution. To evaluate a sample value of $\eta$ with respect to the null distribution we assign it a one-sided $p$-value,
\begin{equation}
p_{\eta} = \int_{\eta}^{\infty} f(x) \ \mathrm{d}x \ = \ \frac{1}{2} \left( 1 + \erf \left(-\frac{\eta}{\sigma \sqrt{2}}\right) \right).
\label{eq:p-value}
\end{equation}

\begin{figure}[!h]
	\centering
		\includegraphics[width=.85\textwidth]{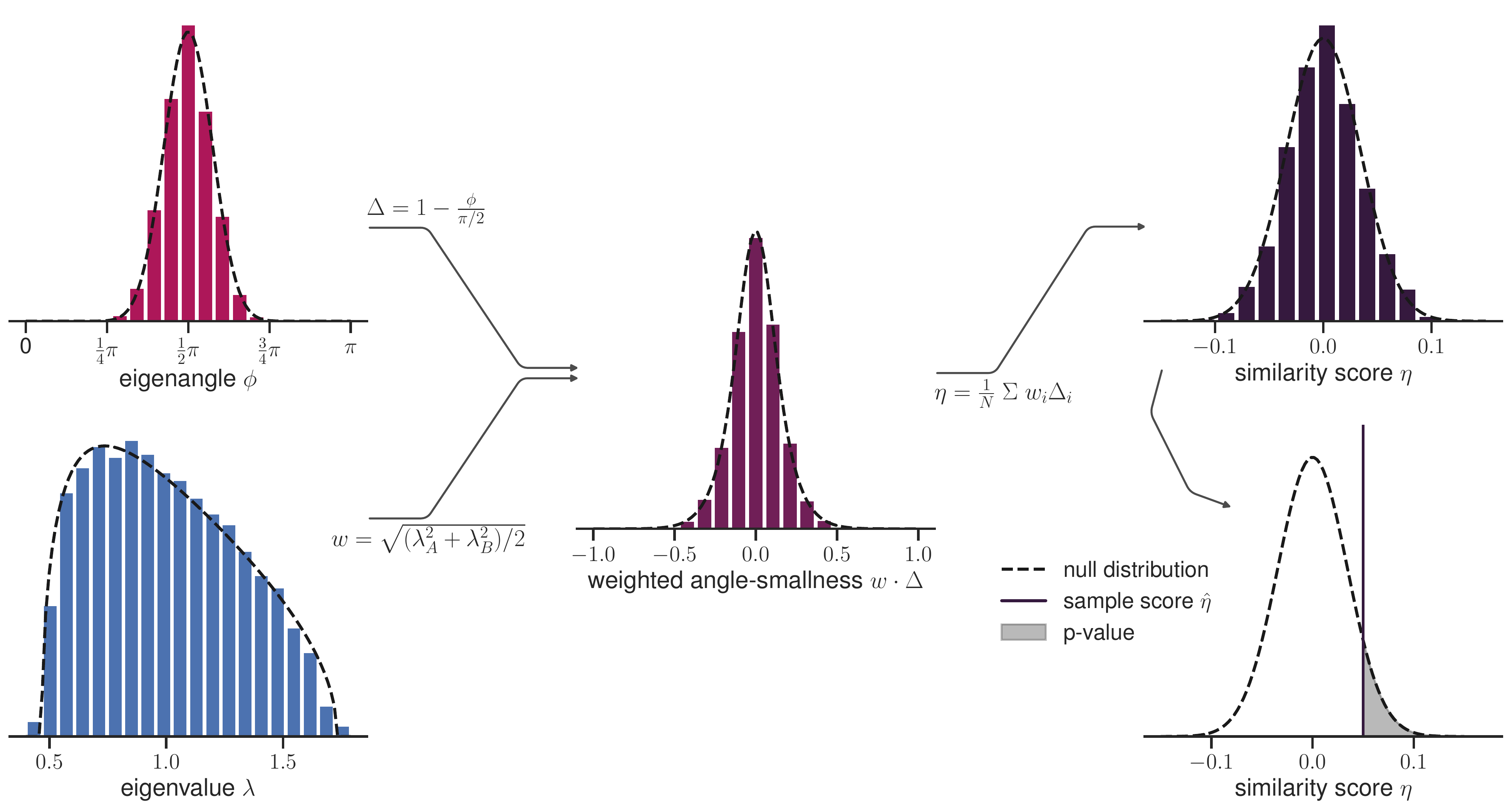}
	  \caption{\textbf{Construction of the eigenangle test.} From left to right the similarity score $\eta$ and its corresponding null distribution is constructed by multiplying the angle-smallness $\Delta$ (derived from the eigenangles $\phi$) with the corresponding eigenvalue weights $w$ (derived from the corresponding eigenvalues $\lambda$) for each pair of matched eigenvectors, and then taking the average. Using the null distribution, $p$-values can be assigned to the similarity score $\eta$. The shown distributions are for dimension $N=20$ and the histograms represent $10^4$ samples.}\label{fig:fig2}
\end{figure}

Using the eigenangle NHST, we conclude the property of interest (here, similarity) as an alternative hypothesis when the null hypothesis (here, independence) is rejected in the case of a $p$-value smaller than a significance level $\alpha$. Often, the significance level is arbitrarily set (e.g., $\alpha=0.05$), which has implications discussed in detail in the literature \citep[e.g.,][]{Nakagawa2007_591a, Szucs2017_}. Alternatively, the $p$-value could be regarded as another quantification, which is a random variable and should be calibrated with suitable reference scenarios. These scenarios should cover different ways in which the underlying null hypothesis could be violated in order to gauge the susceptibility of the test. Therefore, to better asses the eigenangle test, in the following, we explore its behavior in use-cases of generated stochastic activity with inserted correlations (\prettyref{sec:calibration}). Moreover, we extend our test to the case of asymmetric connectivity matrices (\prettyref{sec:asymmetric_extension}) and investigate the influence of synaptic rewiring onto both the connectivity and activity for a simple balanced random network model (\prettyref{sec:rewiring}).

\section{Calibration of eigenangle similarity for correlation structures} \label{sec:calibration}
We motivated the use of angles between pairs of eigenvectors of correlation matrices with the argument that the eigenvectors associated with largest eigenvalues will point to groups of correlated neurons. Coming from this intuition, in the following, we quantify the ability of the presented approach to properly detect correlation structures in neuronal data on the basis of three calibration scenarios.

\begin{figure}[!h]
	\centering
	\includegraphics[width=.85\textwidth]{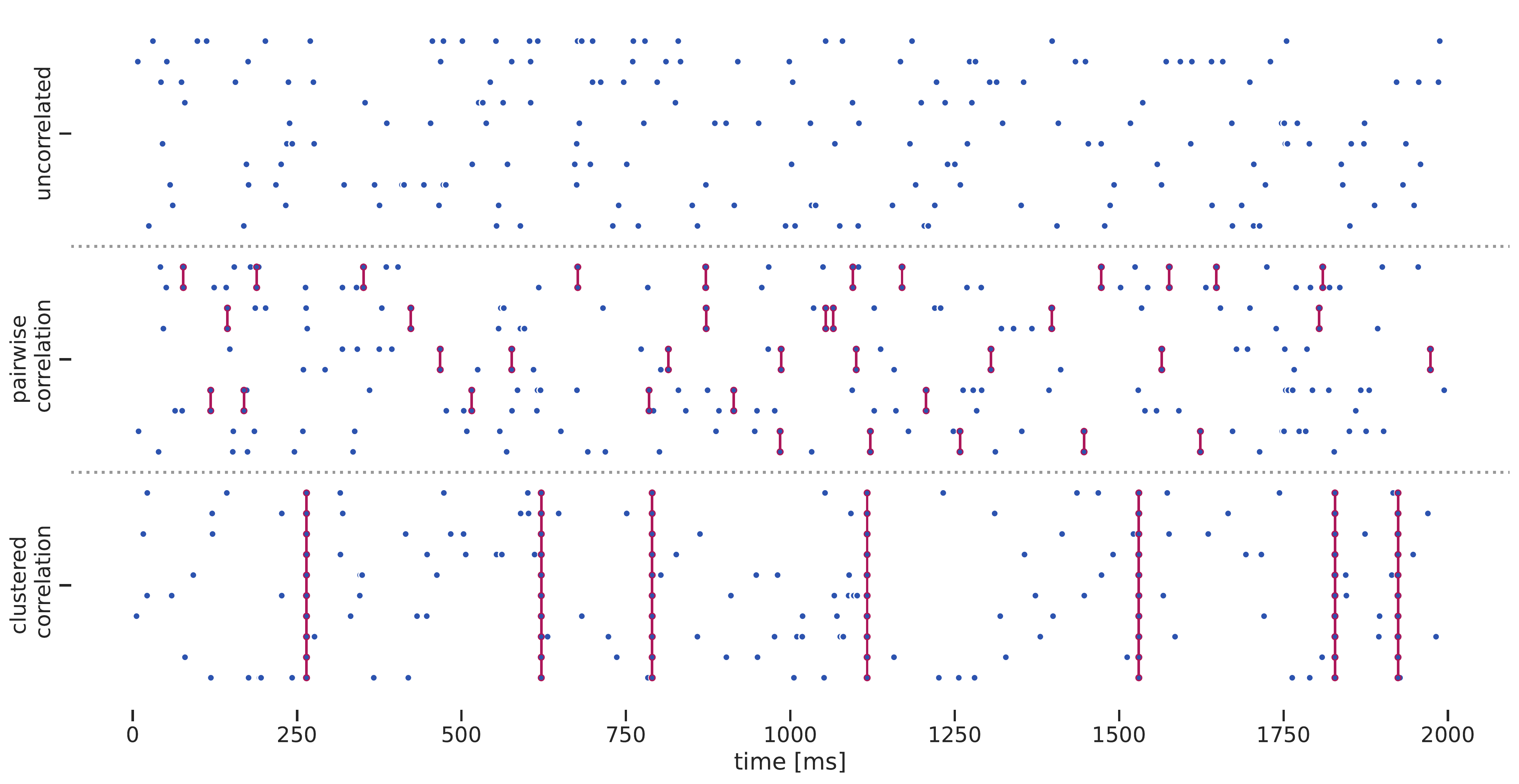}
	\caption{\textbf{Example stochastic spiking activity with inserted correlations}. The rasterplot shows $2$~s of stochastic Poisson spiking activity (rate $10$~Hz) for three correlation scenarios ($10$ spike trains each) created by inserting synchronous spikes (red) with a compound Poisson process. The \textbf{top row} shows independent activity with only chance correlations. The \textbf{middle row} shows activity with synchronous spikes inserted in non-overlapping pairs. The \textbf{bottom row} shows activity with synchronous spikes inserted in across multiple spike trains. The mean correlation coefficient for the pairwise (middle) and clustered (bottom) correlation neurons is $0.3$. \label{fig:stochastic_activity}
	}
\end{figure}

We first create a calibration scenario of two sets of independent stochastic spiking activities consisting of identical numbers of neurons, in which we introduce correlations among the neurons in the same sub-populations, respectively. The bottom row (clustered correlation) of \prettyref{fig:stochastic_activity} shows an example spiking activity of such a correlated sub-population. The larger the strength of these introduced correlations and the size of sub-populations is, the more similar should the test to judge the two correlation matrices and the smaller should the returned $p$-values to be.

\begin{figure}[!h]
	\centering
	\includegraphics[width=.85\textwidth]{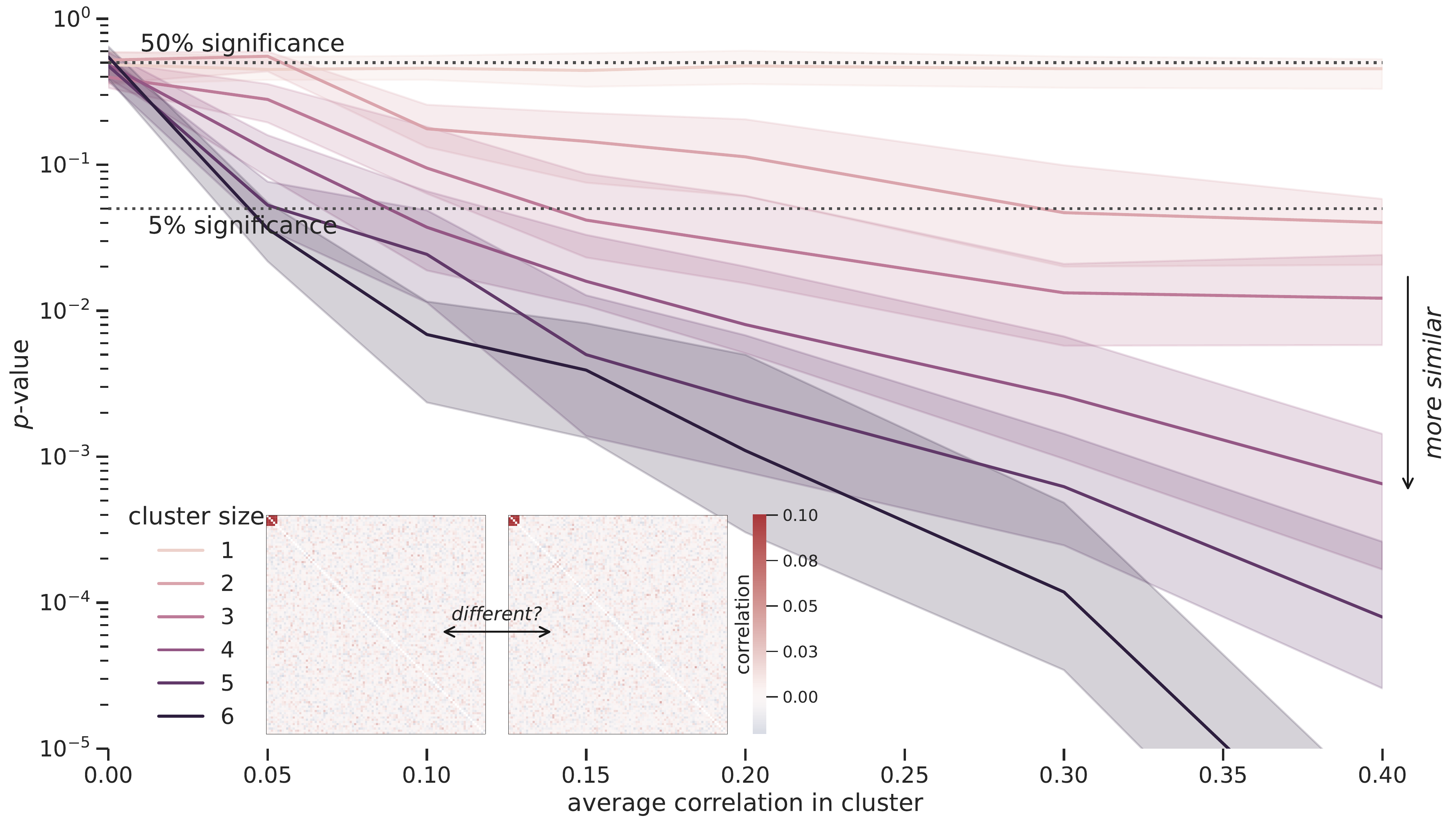}
	\caption{\textbf{Comparing sets of stochastic activity with similar correlation structures}. The resulting $p$-values of the eigenangle test applied to two correlation matrices of stochastic Poisson activity ($100$ neurons) with a respective sub-population (cluster) of given size (color hues) and prescribed average internal correlation. Median (curves) and $95\%$ confidence intervals (shaded areas) of the $p$-value result from $105$ repetitions. The matrices show two representative examples of correlation matrices for simulations with identical parameters.}\label{fig:stochastic_activity_calibration}
\end{figure}

We create a total of $N=100$ spike trains of length $T=30$~s, of which $N-n$ represent independent neurons modeled by a Poisson process (rate $\nu=10$~Hz), and one sub-population (cluster) of size $n$ is modeled by inserting synchronous spikes required to realize a pre-described correlation coefficient via a compound Poisson process \citep{Staude2010_327} that maintains the average rate $\nu$. For the calculation of the correlation matrices we use a bin size of $2$~ms. As \prettyref{fig:stochastic_activity_calibration} shows, the eigenangle test indeed reflects the increasing similarity of the correlation structure with increasing cluster size and cluster correlation. As expected, for zero correlation within the cluster the null hypothesis that the correlation matrices are independent is accepted ($p\sim50$\%). However, for a sufficiently large cluster size and cluster correlation (e.g., size$=4$, correlation $>0.1$) the test rejects the null hypothesis with a given significance level (e.g., $p=5$\%), indicating a similarity of the correlation structures. This analysis confirms the intuition underlying the construction of the eigenangle test and further provides an interpretation of its $p$-value. For example, a significance level of $0.05$ for the eigenangle test corresponds to a shared correlation structure that is as least as dominant as $4\%$ of neurons being correlated with an average coefficient of $0.1$ with the other neurons showing independent activity.

To further characterize the features of the eigenangle test we compare it to the well-established Kolmogorov-Smirnov (KS) test \citep{Hodges1958_469}. As for most common two-sample tests, the null hypothesis of the KS test is that both samples originate from a common underlying distribution, i.e., it tests for similarity and a $p$-value smaller than the significance level rejects this similarity. The eigenangle test, on the other hand, has a null hypothesis regarding the independence of two samples (matrices), i.e., it tests for difference and thus, contrary to the KS and other tests, a significant $p$-value indicates similarity.

When performing the above parameter scan with the KS test applied to the distribution of correlation coefficients there is no visible dependency of the $p$-value on either the cluster size or the correlation and no rejection of the KS null hypothesis with a median $p$-value of $10.4\%$ and $0.5\%$ of all $p$-values smaller than a $1\%$ significance level (Bonferroni corrected). This result is in agreement with the expectation, because in this example the distributions of the correlation coefficients in the two matrices are similar irrespective of the correlation structure. 

\begin{figure}[!h]
	\centering
    \includegraphics[width=.85\textwidth]{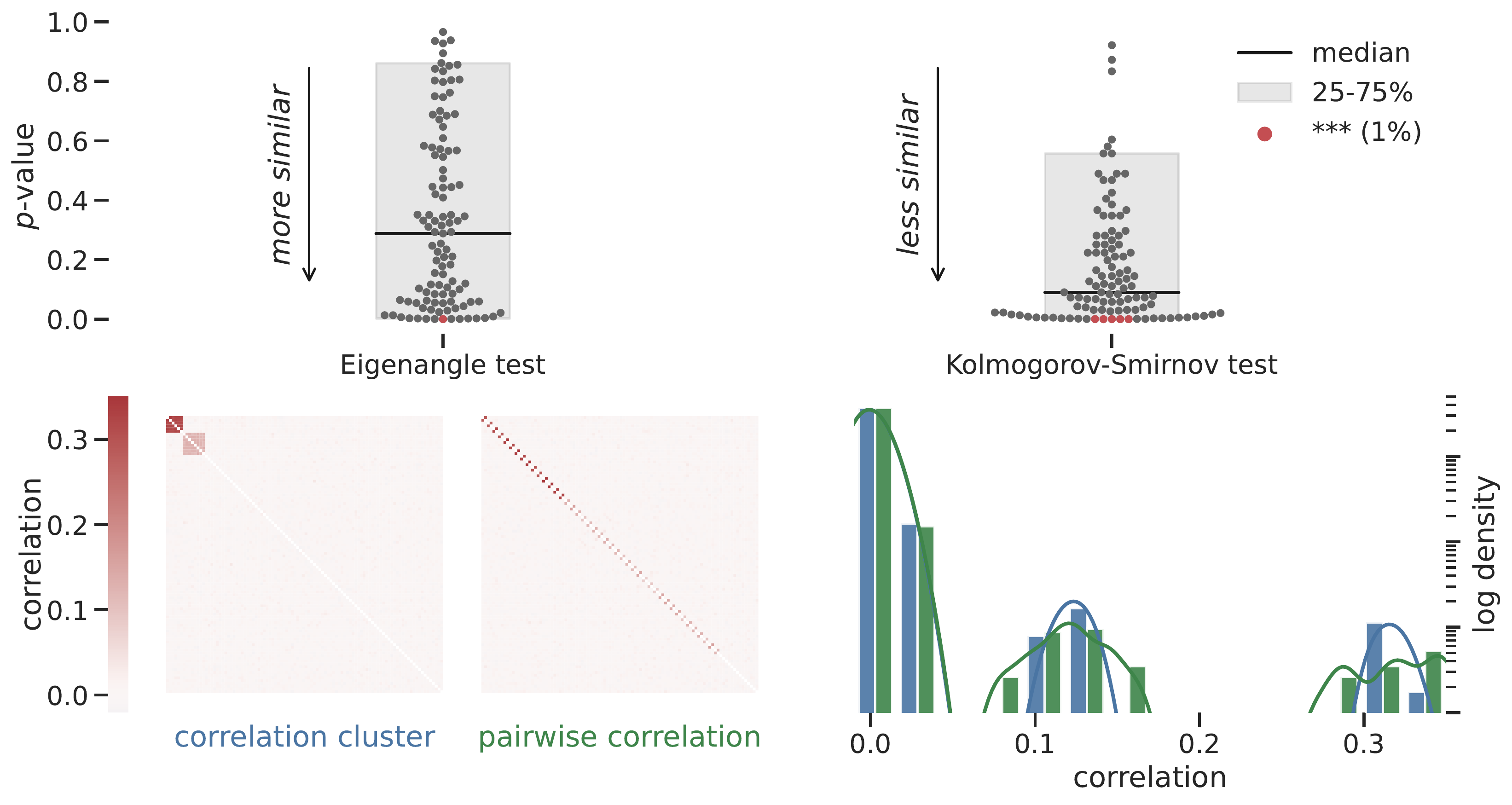}
	\caption{\textbf{Comparing clustered vs. distributed correlations.} Correlations of network activity are organized in distinct groups of multiple neurons or in independent pairs of two. \textbf{Top:} Swarm plots of $p$-values (dots) obtained from the eigenangle (\textit{left}) and KS (\textit{right}) tests for $N=105$ random initializations of clustered and distributed correlations. Red dots: $p$-values significant at the $1\%$ level. \textbf{Bottom:} Example realizations of correlation matrices (\textit{left}) for the clustered (blue) and distributed (green) case, and corresponding distributions of correlation coefficients (\textit{right}) used for the KS test.} \label{fig:CPP_pairwise_comparison}
\end{figure}

For the second calibration scenario, we again generate correlation matrices from stochastic Poisson activity with correlated sub-populations, but now, while one sample has two correlated groups (\prettyref{fig:stochastic_activity}, bottom) of size $6$ and $8$ (average correlation coefficients: $0.3$ and $0.1$, respectively), the other sample has the same "amount" of correlation but distributed among pairs of non-overlapping neurons. The middle row (pairwise correlation) in \prettyref{fig:stochastic_activity} shows an example spiking activity of such correlated neuron pairs. Therefore, the correlation structure of the two samples is distinctly different while the corresponding distributions of correlations coefficients are similar. 
Comparing the matrices with the eigenangle test and the distributions with the KS test illustrates that information about the structure is lost when only comparing the samples of correlation coefficients. \prettyref{fig:CPP_pairwise_comparison} shows that the eigenangle test results in no rejection of the null hypothesis, and thus adequately indicates the difference of the samples with $1.0\%$ false positives for a $1\%$ Bonferroni-corrected significance level ($N=105$). The KS test, in most cases ($95.2\%$), also doesn't reject its null hypothesis and therefore indicates a principal similarity between the distributions as the difference between grouped and pairwise correlation is not represented in the distribution of correlation coefficients. 

In a third calibration scenario, we implement a typical validation use case and compare the correlation matrices from a simulation of the same network implemented on two different network simulation engines (for details, see \prettyref{suppfig:simulator_comparison}). In this scenario, we expect that both simulations exhibit activity with a matching correlation structure, given that the underlying structure of connectivity in the network model as well as the input is identical. Indeed, the null hypotheses for the eigenangle test is rejected, indicating that the two simulations produce a similar structure of correlation. However, as the null hypothesis for the KS test is rejected as well, this indicates that there are different distributions of correlation coefficients.

The three calibration scenarios illustrate not only the proper behavior of the eigenangle test, but also how it is complementary to classical two-sample tests in explicitly evaluating structural matrix features above the distribution of values.

\section{Statistical eigenangle test for connectivity matrices} 
\label{sec:asymmetric_extension}
In the following, we investigate under which conditions it is possible to go beyond the description of the eigenangle score for correlation matrices and adapt the statistical test to other kinds of matrices representing pairwise measures. In fact, the null distribution for the eigenangle score can be formulated for any kind of random matrix given two requirements: they have an analytic description for their eigenvalue distribution, and the angles between corresponding eigenvector pairs of two independent realizations of the matrix are distributed like random angles. 
This opens up additional applications of this approach, in particular the extension to connectivity matrices that we discuss in the following. Random connectivity matrices differ considerably from random correlation matrices, for example, in that they are sparse, may be inherently structured by the connectivity parameters of different sub-populations of neurons, and have complex-valued eigenvalues and eigenvectors since they are typically asymmetric.

\begin{figure}[!h]
	\centering
      \includegraphics[width=.8\textwidth]{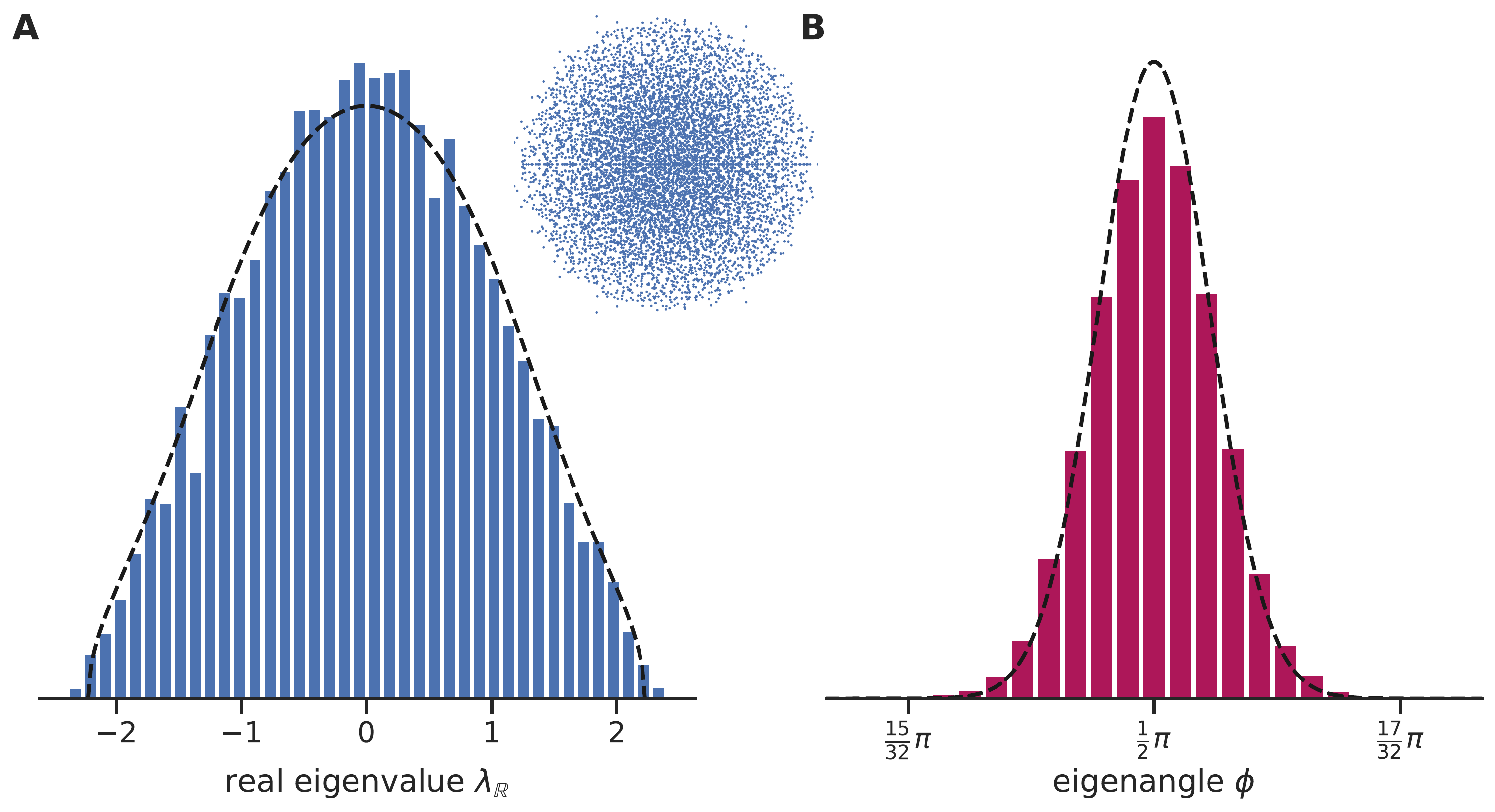}
	  \caption{\textbf{Eigenvalue and eigenangle distribution of connectivity matrices.} For the connectivity matrix of a random balanced network (seed details in \prettyref{tab:network_configuration}), panel \textbf{A} shows the sampled distribution of real eigenvalues (inset showing the scatter plot of complex eigenvalues) and the adapted Rajan-Abbott distribution. \textbf{B)} The sampled eigenangle distribution of the corresponding complex-valued eigenvectors and the prediction for random angles $\propto \sin\phi^{2N-2}$. The sampled histograms are based on eight re-initializations of the network's connectivity.
	  }
	  \label{fig:connectivity_dists}
\end{figure}

We consider here a specific type of network that fulfils the above two assumptions. \citet{Rajan2006_188104} present an analytic description for the absolute eigenvalue values of the connectivity matrices of balanced excitatory-inhibitory (E-I) networks consisting of two (potentially sparsely connected) sub-populations with their respective synaptic weight distributions. The connections are drawn independently without adhering to a fixed in- our out-degree (i.e., 'pairwise Bernoulli'). In this study, for simplicity we define these two sub-populations to consist of the E and I neurons, respectively. This implies two weight distributions for the E and I connections. To rank the eigenvectors and set the weights $w_i$ for the eigenangles we use only the real component $\lambda_\mathbb{R}$ of the eigenvalues. The corresponding distribution $h(\lambda_{\mathbb{R}})$ can be derived from the distribution of the absolute values of the eigenvalues $p(|\lambda|)$ by exploiting the point-symmetry of the spectrum for independently drawn matrix elements \citep{Sommers1988_1895, Girko1985_694},
\begin{equation}
    h(\lambda_\mathbb{R}) = \int_{|\lambda_\mathbb{R}|}^{r_c} \frac{r}{\sqrt{r^2-\lambda_\mathbb{R}^2}} p(r) \ \mathrm{d}r,
\end{equation}
where $r_c$ is the radius of the circle of complex eigenvalues. In the following, we are working with a specific network model incorporating the above properties. The exact configuration of the network model is documented in \prettyref{tab:network_configuration}. Using random initialization of this network model, \prettyref{fig:connectivity_dists}A displays the sampled histogram the real eigenvalues together with the analytical description.

We further observe that when the total weight variances of the two populations (taking into account also the variance caused by their relative size difference) are approximately the same, the eigenangles are well described by the distribution of random angles. As the eigenvectors ($\mathbf{v}$) are now complex-valued we just need to adapt the definition of angles to $\phi=\arccos{(<\mathbf{v}_A, \mathbf{v}_B^*>_{\mathbb{R}})}$ with $<,>$ representing the inner product and $^*$ the complex conjugate. Since the additional imaginary vector components double the degrees of freedom, the dimensionality factor $N$ in \prettyref{eq:angle-distribution} also needs to be adapted accordingly to $2N$ so that the random eigenangle distribution becomes
\begin{equation}
f_\measuredangle(\phi) = \frac{\Gamma(N)}{\sqrt{\pi} \Gamma(N-\frac{1}{2})} \sin(\phi)^{2N-2} \qquad \phi \in [0,\pi].
\label{eq:angle-distribution-connectivity}
\end{equation}

\prettyref{fig:connectivity_dists}B displays this analytical eigenangle distribution and the histogram of sampled eigenangles for the comparison of re-initializations of our network model. Here, the weight variances in the network are not exactly equal but $\frac{\mathrm{Var} \ J_\mathrm{I}}{\mathrm{Var} \ J_\mathrm{E}} = 6.16 > 1$. This deviation of the weight variance ratio from $1$ causes a slightly mismatch of the sampled eigenangle distribution with the analytic curve. However, further adjusting the weight variances would require the standard deviation of the underlying lognormal weight distribution to become considerably larger than its mean, leading to a worse estimation of the sample variance of the weights and consequently a worse fit of the eigenvalue distribution. Hence, the chosen configuration represents a compromise for this kind of E-I networks with an E/I ratio of $80\%$ to $20\%$ to still allow for a reasonable fit of the eigenangle score distribution.

Using this extension to connectivity matrices, we are able to quantitatively compare both the connectivity and the activity correlations of certain classes of neural networks with the same statistical method and are therefore able to directly relate the changes in one to the changes in the other. This provides a unique approach to investigate to what degree features of the network connectivity determine aspects of the neural activity.

\section{Evaluating network rewiring effects on the correlation- and connectivity structure} \label{sec:rewiring}
Using the extension of the eigenangle score for the connectivity matrix of the specific network type described in \prettyref{sec:asymmetric_extension}, we evaluate the effect that a rewiring of the synaptic couplings of a random balanced network model has on the correlation structure of the simulated activity. These types of simulation experiments are inspired by \citet{Mongillo2018_1463} who evaluated the strategic shuffling and adding of synapses to explore the effects of synaptic volatility and learning. In their study, the authors quantified the influence of modifying connectivity on the activity by using the Pearson correlation between the firing rate vectors before and after rewiring and by calculating the element-by-element correlation of the connectivity matrices. In the following, we demonstrate how the eigenangle test applies to this scenario and how it can be used to quantify the relationship between structural and functional changes in the network. 

\begin{figure}[!h]
	\centering
	\includegraphics[width=.95\textwidth]{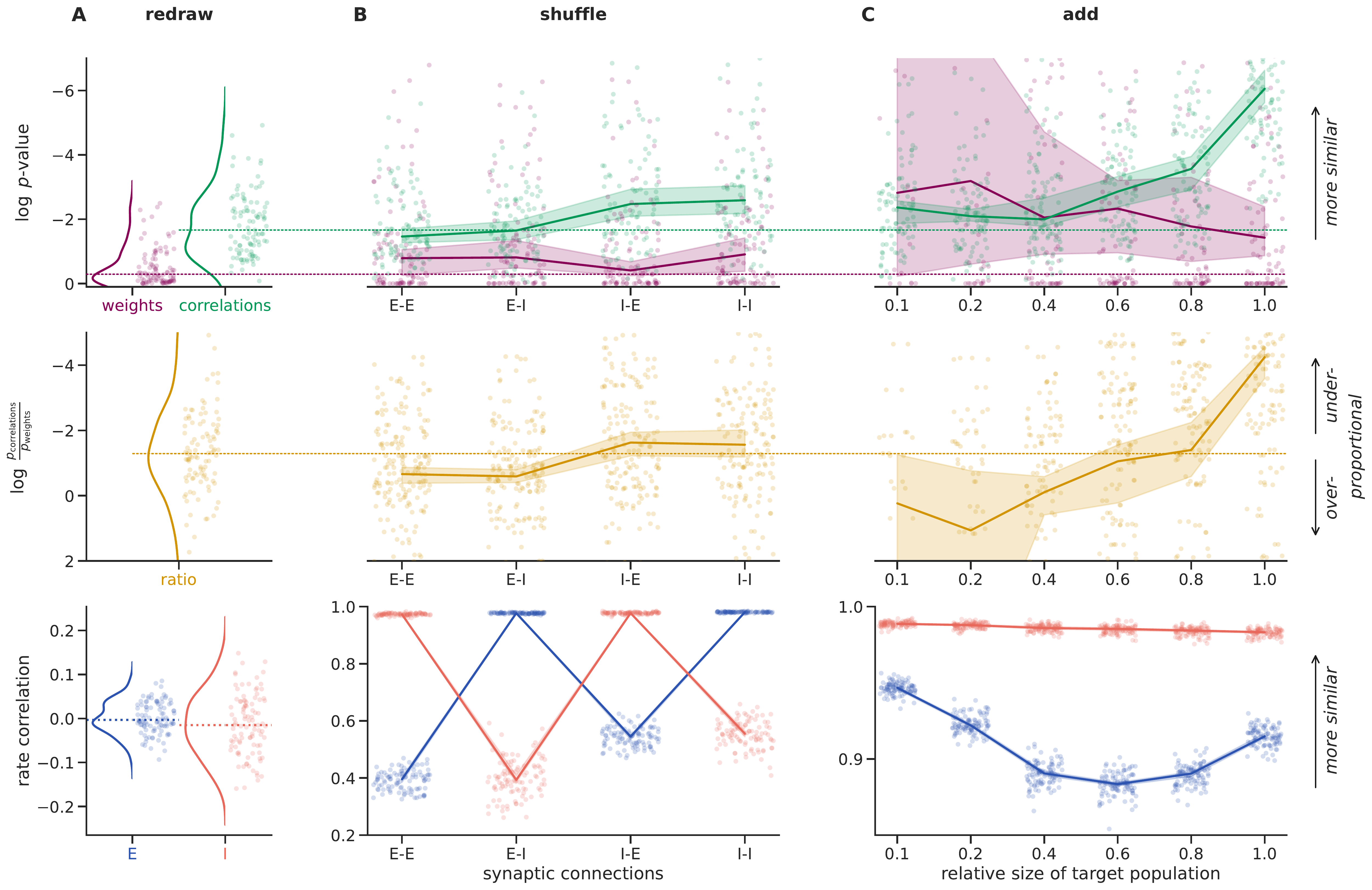}
	\caption{\textbf{Comparing the effects of rewiring protocols on the connectivity and activity of a random network model.} In each panel, the scattered points show the outcomes of $100$ repeated comparisons between original and rewired networks realizations (some points lie outside the bounds of the plotted domain and are not shown), and the curve and shaded area show the median value and the bootstrapped $95\%$ confidence interval. \textbf{Top row:} log $p$-values of the eigenangle test comparing the weight (magenta) and correlation (green) matrices of two network realizations. \textbf{Middle row:}  corresponding ratios of the log $p$-values for the comparisons of correlations versus corresponding weights indicate if changes in activity correlation are over- or under-proportional given the change in the connectivity structure. \textbf{Bottom row:} correlation of the vector of firing rates, separated for excitatory (E, blue) and inhibitory (I, red) neurons. \textbf{A)} The "redraw" column shows the distribution of comparison results when comparing two random network initialization, i.e., different random seeds. \textbf{B)} The "shuffle" column compares a networks with rewired versions where the synapses between the populations indicated on the axis are shuffled. \textbf{C)} The "add" column compares networks with rewired versions where the number of E-E synapses, that targeted a sub-population constituting a relative fraction of $x$ of the target E population, is increased by $20\%$.} \label{fig:rewiring_comparison}
\end{figure}

For these simulation experiments we use the network model introduced in the previous Section, implemented with the Nest simulator\footnote{RRID:SCR\_002963} \citep{Nest3.1}. The network activity is in an asynchronous state, but exhibits isolated periods of synchronization between neurons (\prettyref{suppfig:rasterplot}).
Based on this network, in analogy to the work by \citet{Mongillo2018_1463}, we analyze three types of rewiring protocols.
\begin{itemize}
    \item[] \textbf{Redraw:} the entire network is re-initialized with a different random seed, i.e., a new connectivity matrix is drawn from the identical weight distribution as the original network.
    \item[] \textbf{Shuffle:} keeping the exact weight values, the synaptic weights between all pairs of neurons in between a source and a target population are shuffled,
    \item[] \textbf{Add:} new E-E synapses are drawn that target a sub-group constituting a fraction $x$ of the excitatory population. Synaptic weights are sampled from the weight distribution of existing E-E connections. The number of new synapses is $20\%$ of the number of existing synapses towards the target sub-group, i.e., $0.2 x$ of all E-E synapses.
\end{itemize}
The results of performing each of the comparisons (original vs.\ rewired network) for each protocol 100 times with different random initialization is illustrated in \prettyref{fig:rewiring_comparison}. \\

\textit{Redraw -}
The redrawing protocol represents a scenario not unlike the null hypothesis of the eigenangle test, in the sense that two independent random connectivity matrices are compared that share the statistical properties. As expected, the eigenangle test results in non-significant $p$-values (with a median of $p=0.52$) for the comparison of the synaptic weights (\prettyref{fig:rewiring_comparison}A, top). The comparison of the resulting correlation matrices, on the other hand, indicates a certain degree of similarity (with a median of $p=0.02$). This initialization-independent similarity between the correlation structures of completely rewired networks may be explained by the influence of the general network configuration and state of the population dynamics. Intuitively, networks that exhibit a stronger fluctuation of the population activity and therefore a stronger correlation between many neurons are consequently less effected by the redrawing protocol. This characterization of the network's susceptibility to synaptic changes provides a reference to contextualize the effect of the other rewiring protocols. \\

\textit{Shuffle -}
\prettyref{fig:rewiring_comparison}B illustrates the effects of shuffling the existing synaptic connections between and within the E and I populations (i.e., E-E, E-I, I-E, and I-I). The shuffling of the synaptic weights from E neurons leads to a larger change (i.e., less similarity) in the overall correlation structure than the shuffling of weights where the source population is I. The changes to the structure of the weight matrix are, however, rather similar for all combinations of source and target populations, and only show slightly larger $p$-values for the E-E shuffling than the I-I shuffling. This can be attributed to the larger number of synapses that were shuffled. Relating the trends of the $p$-value of correlation and weight comparison by means of the ratio ($\log (p_\mathrm{correlations}/p_\mathrm{weights})$) we may gauge whether the changes to the correlation structure are under- or over-proportional as compared to the changes to the connectivity structure (\prettyref{fig:rewiring_comparison}B, middle), especially when we compare the ratio to the reference scenario of a complete redrawing of the weights (\prettyref{fig:rewiring_comparison}A, middle). This way, we identify an over-proportional influence of changing synapses originating from the E population on the correlation structure.

Additionally, we show the effects of shuffling using the Pearson correlation coefficients of the E and I rate vectors respectively before and after rewiring (\prettyref{fig:rewiring_comparison}B, bottom). The shuffling only has an effect on the firing rate vector of the target population, whereas the vector of the source population is nearly perfectly correlated, i.e., unchanged by the rewiring protocol. Further, the plot shows a larger influence of synapses originating from the E population onto the rates of the target population as compared to those originating from the I population. However, this trend inverts when correcting for the difference in the total number of synapses in the E and I populations. When shuffling the same number of weights in each case, the synapses originating in the I population exhibit the stronger influence onto the rate vectors as compared to those originating the in E population, in particular for I-I connections (\prettyref{suppfig:rewire_comparisons_synapse_corrected}B, bottom). However, the trend that E synapses have greater influence on the correlation structure as measured by the eigenangle test remains even with the correction for the number of synapses (see \prettyref{suppfig:rewire_comparisons_synapse_corrected}B, top and middle).

Similar to our simulation experiment, \citet{Mongillo2018_1463} also show a larger influence of the shuffling on the rates of the targeted population, but in contrast also show a considerable influence on the non-targeted population. Furthermore, the study reports a greater influence of synapses originating in the I population even when not correcting for the number of shuffled synapses. These deviations of the results of a shuffling protocol emphasize the relevance of the underlying network configuration. In contrast to the network employed in this study, \citeauthor{Mongillo2018_1463} investigate a network that contains considerably higher connection probabilities (and thus, more indirect and recurrent connections), exhibits a variance of the I weight distributions that exceeds that of the E weight distributions, and variance of the I firing rates that exceeds that of the E firing rates. The latter two factors are highlighted by the authors as the relevant attributes for the increased influence of I connections. \\

\textit{Add -}
In the third rewiring protocol (\prettyref{fig:rewiring_comparison}C), as increasingly more E-E synapses are added with an increasing target fraction $x$, the similarity between the original and the rewired correlation structure also increases. This indicates that additional synapses have an influence on the correlation structure that becomes larger when synapses are targeted towards a smaller sub-population, even when fewer new synapses are added as for larger target sub-populations. With the increase of $20\%$ of synapses targeting the full E population at $x=1.0$ (equivalent to $12800$ synapses) the correlation structure remains more similar than when increasing the number of synapses targeting $x=0.1$ of the E population by $20\%$ (equivalent to $1280$ synapses). The influence of the latter intervention is even close to the effect of a complete rewiring of the network.

With the increase in total E-E connections there is more and more synchronous network activity introduced. Still, even though the average correlation becomes considerably larger, when the new synapses are distributed randomly over the E population the tests detects a distinctly similar correlation structure. This illustrates how the test is indeed complementary to classical two-sample tests, focusing on the precise organisation of correlations instead of the amount (i.e., its sample distribution).

The influence of adding synapses on the similarity of the weight matrix seems to be linked to the total number of added synapses, i.e., more similar the fewer synapses are added. The same trend of the weight matrix similarity is also shown in \citet{Mongillo2018_1463} by means of the correlation coefficient between the original and rewired matrix.
Combining the trends of the correlations and weights we see the over-proportional effect of adding a few targeted synapses, while adding many distributed synapses tends to have an under-proportional effect.
\citeauthor{Mongillo2018_1463} arrive at a similar conclusion for the rate correlations, which, in their study, are close to $1$ for both extremes of the protocol, the new synapses being distributed throughout the entire population and diminishingly few synapses target onto a progressively negligible sub-population. They report the maximal influence, i.e., the minimum of the resulting U-shaped rate correlation curve, at $0.2$. In our simulations, we also find the U-shape effect in the rate correlation. However, as already observed with the shuffle protocol, the addition of E-E connections has only effect on rates of the immediate targets and the rates of the I neurons are basically unchanged. Further, we find the maximal influence on the E rate correlation for a target population of $0.6$, indicating that the trade-off between the vanishing effects of too few and too distributed additional synapses strongly depends on the network configuration. 
When changing the protocol and adding the same number of synapses irrespective of the target population size the isolated effect of focusing the new connections becomes more visible (see \prettyref{suppfig:rewire_comparisons_synapse_corrected}). The influence on the rate correlations, in particular for E, is stronger for smaller target populations. The corresponding trend for the correlation structure remains approximately the same with this protocol, except that the range of increasing influence is for target populations from $0.4$ to $0.1$, whereas with the prior add protocol this range of increasing influence is from $1.0$ to $0.4$ before saturating at the baseline for $<0.4$. As noted before, the influence on the structure of the weight matrix is dominated by the total number of new synapses, therefore when adding the same number of synapses of each target size the trend curve stays flat.


\section{Conclusion}\label{sec:conclusion}
Angles quantify the alignment between vectors. We show that the angles between the eigenvectors of two matrices weighted by the corresponding eigenvalues provide a means to quantify the similarity between matrices. Further, we derive the probability distribution of the resulting similarity score for the null hypothesis of independence for random correlation matrices and a class of random connectivity matrices. From these components, we define the statistical "eigenangle" test to evaluate shared non-random structures in two matrices, in particular with application to correlation and connectivity matrices of neural networks. 

The eigenangle test expands a niche of statistically comparing pairwise measures, which can be represented as matrices. While classical two-sample tests (like the Kolmogorov-Smirnov test) compare sample distributions, the matrix comparison retains the relations between the pairwise values and can therefore evaluate the structure, and not only the amount of, for example, correlations in a network.

We demonstrate the behavior of the eigenangle test in three calibration scenarios, showing that it properly measures shared structural features (i.e., a correlated sub-group) and distinguishes different structural arrangements even when the distribution of values is similar.

Using a random balanced network model, we employ the eigenangle test to explore the effects of applying different synaptic rewiring protocols on both the synaptic weights and the correlation matrix by quantifying the degree of similarity of the matrices before and after the rewiring. Measuring the magnitude of the observed effects enables to gauge which changes in the connectivity elicit an over- or under-proportional change in the correlation structure.
We find that the kind of change and the magnitude of change caused by a specific synaptic rewiring in the network strongly depends on the exact network configuration, so that it is ambitious to deduce general statements about the functional properties of random balanced networks. However, with our network model we find that the activity correlation structure is mainly determined by the synapses originating in the E population, and that it is more susceptible to small but targeted rewirings as compared to larger but randomly distributed ones. In contrast, the firing rates are more affected by the rewiring of synapses originating in the I population. The effect on the structure of the weight matrix seems, however, to be rather determined by the number of changed synapses (relative to the total number synapses in the respective source-target population) than their organization.

While these results for the rewiring protocols of shuffling and adding synapses, inspired by \citet{Mongillo2018_1463}, mostly confirm the trends observed in that publication and extend them to utilize the proposed eigenangle test to compare connectivity and correlation structures, they also show some discrepancies. These discrepancies are likely explained by key differences in the network models analyzed in our studies, like their connection probabilities and synaptic weight variances. We note that the network model analyzed by \citet{Mongillo2018_1463} does not belong to the class of network models satisfying the conditions required to apply the eigenangle test.

Nevertheless, the eigenangle test offers the application to a wide range of further rewiring protocols to probe the influence of distinct features of network connectivity. Examples of such rewiring protocols could address the arrangement, composition, and interaction of cell assemblies. Many studies use some sort of assemblies in their theories and models \citep{Harris2005_399, Litwin-Kumar2012_1498, Aviel2003_1321}, details of which could be evaluated by means of the eigenangle similarity. 

The application of the test is restricted to networks with a large number of neurons as a result of the assumptions underlying the null hypothesis. We estimate, however, on the basis of numerical simulations that for practical purposes $N \ge 100$ represents a sufficiently large network size. Further, the networks to be compared need to have the same neuron identities to be able to properly define the corresponding eigenangles. Thus, one valid application is the comparison of two versions of the same network model. Besides the rewiring experiments discussed above, this also includes validation testing of the sort evaluating the model's dependency on the input, its robustness to parameter variations, or its stability over time. Another potential application for the eigenangle test is the comparison of recordings of neural activity in which identical neurons were identified. Hence, the test can quantify the similarity of the correlation structure across different task conditions or, for example, before and after a stimulus presentation. In all usage scenarios, calibrating the test's behavior with respect to various modelled influences on connectivity and correlation, as shown in \prettyref{sec:calibration}, can assist in interpreting the results. Importantly, classical validation of a model against an experiment is in general not possible, unless a mapping between simulated and experimentally observed neurons can be made.

Our results emphasize that the influence of connectivity on the neural dynamics can be complex and non-linear as, for example, also shown in \citet{Dahmen2022_e68422, Wainrib2013_118101, GarciadelMolino2013_042824, Sompolinsky1988_259}. However, we show that there are strategic approaches to evaluate these aspects for a given neural network dynamics. In analogy to how we adopted the eigenangle test to the connectivity of network types described by \citet{Rajan2006_188104}, the test can in principle be further expanded with the analytical eigenspectrum descriptions of the connectivity of other network types \citep{Sommers1988_1895, Kuczala2016_050101, Ahmadian2015_012820, Schuessler2020_013111, Muir2015_042808}. Further, one might also define more specific null hypotheses than having random correlations by determining the null distribution numerically via surrogate data \citep{Grun2009_1126, Stella2021_2021.08.24.457480}.
With this outlook, and building on previous work that explores the informative value of the connectivity eigenspectrum onto the network state and dynamics \citep{Zhou2009_2931, Dahmen2019_13051}, we suggest that the concrete quantification in form of the eigenangle similarity score and a probability value can support the numerical evaluation and understanding of various kinds of network models.

\section*{Acknowledgements}
This project has received funding from the European Union’s Horizon 2020 Framework Programme for Research and Innovation under Specific Grant Agreements No. 785907 (Human Brain Project SGA2) and No. 945539 (Human Brain Project SGA3), the Helmholtz Association Initiative and Networking Fund under project number ZT-I-0003, and the Deutsche Forschungsgemeinschaft (DFG, German Research Foundation) – 491111487.

We thank Alexander van Meegen for his valuable insights on neural network dynamics.


\section*{CRediT authorship contribution statement}
\textbf{Robin Gutzen:} Conceptualization, Software, Methodology, Writing - Original Draft, Visualization. \textbf{Sonja Grün:} Writing - Review \& Editing; Supervision; Funding acquisition. \textbf{Michael Denker:} Conceptualization, Writing - Review \& Editing, Supervision; Funding acquisition.

\bibliography{eigenangles}

\begin{thebibliography}{}
\expandafter\ifx\csname natexlab\endcsname\relax\def\natexlab#1{#1}\fi
\providecommand{\url}[1]{\href{#1}{#1}}
\providecommand{\dodoi}[1]{doi:~\href{http://doi.org/#1}{\nolinkurl{#1}}}
\providecommand{\doeprint}[1]{\href{http://ascl.net/#1}{\nolinkurl{http://ascl.net/#1}}}
\providecommand{\doarXiv}[1]{\href{https://arxiv.org/abs/#1}{\nolinkurl{https://arxiv.org/abs/#1}}}

\bibitem[{Ahmadian {et~al.}(2015)Ahmadian, Fumarola, \&
  Miller}]{Ahmadian2015_012820}
Ahmadian, Y., Fumarola, F., \& Miller, K.~D. 2015, Physical Review E, 91,
  012820, \dodoi{10.1103/PhysRevE.91.012820}

\bibitem[{Albert {et~al.}(2002)Albert, Barab{\'a}si, Barabasi, \&
  Barab{\'a}si}]{Albert2002_47}
Albert, R. R.~R., Barab{\'a}si, A.-L., Barabasi, A.-L., \& Barab{\'a}si, A.-L.
  2002, Reviews of Modern Physics, 74, 47, \dodoi{10.1103/RevModPhys.74.47}

\bibitem[{Aviel {et~al.}(2003)Aviel, Mehring, Abeles, \& Horn}]{Aviel2003_1321}
Aviel, Y., Mehring, C., Abeles, M., \& Horn, D. 2003, Neural Computation, 15,
  1321, \dodoi{10.1162/089976603321780290}

\bibitem[{Bullmore \& Sporns(2009)}]{Bullmore2009_186}
Bullmore, E., \& Sporns, O. 2009, Nature Reviews Neuroscience, 10, 186,
  \dodoi{10.1038/nrn2575}

\bibitem[{Cai {et~al.}(2013)Cai, Fan, \& Jiang}]{Cai2013_1837}
Cai, T.~T., Fan, J., \& Jiang, T. 2013, Journal of Machine Learning Research,
  14, 1837

\bibitem[{Calsbeek \& Goodnight(2009)}]{Calsbeek2009_2627}
Calsbeek, B., \& Goodnight, C.~J. 2009, Evolution, 63, 2627,
  \dodoi{10.1111/j.1558-5646.2009.00735.x}

\bibitem[{Cohen(1988)}]{Cohen1988_a}
Cohen, J. 1988, Statistical {{Power Analysis}} for the {{Behavioral Sciences}},
  2nd edn. ({New York}: {Routledge}), \dodoi{10.4324/9780203771587}

\bibitem[{Curto \& Morrison(2019)}]{Curto2019_11}
Curto, C., \& Morrison, K. 2019, Current Opinion in Neurobiology, 58, 11,
  \dodoi{10.1016/j.conb.2019.06.003}

\bibitem[{Dahmen {et~al.}(2019)Dahmen, Gr{\"u}n, Diesmann, \&
  Helias}]{Dahmen2019_13051}
Dahmen, D., Gr{\"u}n, S., Diesmann, M., \& Helias, M. 2019, Proceedings of the
  National Academy of Sciences, 116, 13051, \dodoi{10.1073/pnas.1818972116}

\bibitem[{Dahmen {et~al.}(2022)Dahmen, Layer, Deutz, D{\k{a}}browska, Voges,
  {von Papen}, Brochier, Riehle, Diesmann, Gr{\"u}n, \&
  Helias}]{Dahmen2022_e68422}
Dahmen, D., Layer, M., Deutz, L., {et~al.} 2022, eLife, 11, e68422,
  \dodoi{10.7554/eLife.68422}

\bibitem[{Dasbach {et~al.}(2021)Dasbach, Tetzlaff, Diesmann, \&
  Senk}]{Dasbach2021_757790}
Dasbach, S., Tetzlaff, T., Diesmann, M., \& Senk, J. 2021, Frontiers in
  Neuroscience, 15, 757790, \dodoi{10.3389/fnins.2021.757790}

\bibitem[{Deepu {et~al.}(2021)Deepu, Spreizer, Trensch, Terhorst, Vennemo,
  Mitchell, Linssen, M{\o}rk, Morrison, Eppler, Kamiji, {de Schepper},
  Kitayama, Kurth, {Morales-Gregorio}, Nagendra~Babu, \& Plesser}]{Nest3.1}
Deepu, R., Spreizer, S., Trensch, G., {et~al.} 2021, {{NEST}} 3.1, Zenodo,
  \dodoi{10.5281/zenodo.5508805}

\bibitem[{Flury(1988)}]{Flury1988_}
Flury, B. 1988, Common Principal Components \& Related Multivariate Models
  ({New York}: {John Wiley \& Sons, Inc})

\bibitem[{{Garc{\'i}a del Molino} {et~al.}(2013){Garc{\'i}a del Molino},
  Pakdaman, Touboul, \& Wainrib}]{GarciadelMolino2013_042824}
{Garc{\'i}a del Molino}, L.~C., Pakdaman, K., Touboul, J., \& Wainrib, G. 2013,
  Physical Review E, 88, 042824, \dodoi{10.1103/PhysRevE.88.042824}

\bibitem[{Girko(1985)}]{Girko1985_694}
Girko, V.~L. 1985, Theory of Probability \& Its Applications, 29, 694,
  \dodoi{10.1137/1129095}

\bibitem[{Gr{\"u}n(2009)}]{Grun2009_1126}
Gr{\"u}n, S. 2009, Journal of Neurophysiology, 101, 1126,
  \dodoi{10.1152/jn.00093.2008}

\bibitem[{Guhr {et~al.}(1998)Guhr, {M{\"u}ller{\textendash}Groeling}, \&
  Weidenm{\"u}ller}]{Guhr1998_189}
Guhr, T., {M{\"u}ller{\textendash}Groeling}, A., \& Weidenm{\"u}ller, H.~A.
  1998, Physics Reports, 299, 189, \dodoi{10.1016/S0370-1573(97)00088-4}

\bibitem[{Gutzen {et~al.}(2018)Gutzen, {von Papen}, Trensch, Quaglio, Gr{\"u}n,
  \& Denker}]{Gutzen2018_90}
Gutzen, R., {von Papen}, M., Trensch, G., {et~al.} 2018, Frontiers in
  Neuroinformatics, 12, 90, \dodoi{10.3389/fninf.2018.00090}

\bibitem[{Haber \& Schneidman(2020)}]{Haber2020_2020.04.27.057752}
Haber, A., \& Schneidman, E. 2020, bioRxiv, 2020.04.27.057752,
  \dodoi{10.1101/2020.04.27.057752}

\bibitem[{Harris(2005)}]{Harris2005_399}
Harris, K.~D. 2005, Nature Reviews Neuroscience, 6, 399,
  \dodoi{10.1038/nrn1669}

\bibitem[{Hodges(1958)}]{Hodges1958_469}
Hodges, J.~L. 1958, Arkiv f\"or Matematik, 3, 469, \dodoi{10.1007/BF02589501}

\bibitem[{Holmes(1991)}]{Holmes1991_239}
Holmes, R.~B. 1991, SIAM journal on matrix analysis and applications, 12, 239,
  \dodoi{10.1137/0612019}

\bibitem[{Izhikevich(2006)}]{Izhikevich2006_245}
Izhikevich, E.~M. 2006, Neural Computation, 18, 245,
  \dodoi{10.1162/089976606775093882}

\bibitem[{Krzanowski(1990)}]{Krzanowski1990_81}
Krzanowski, W.~J. 1990, Journal of Classification, 7, 81,
  \dodoi{10.1007/BF01889705}

\bibitem[{Kuczala \& Sharpee(2016)}]{Kuczala2016_050101}
Kuczala, A., \& Sharpee, T.~O. 2016, Physical Review E, 94, 050101,
  \dodoi{10.1103/PhysRevE.94.050101}

\bibitem[{Kullback \& Leibler(1951)}]{Kullback1951_79}
Kullback, S., \& Leibler, R.~A. 1951, The Annals of Mathematical Statistics,
  22, 79, \dodoi{10.1214/aoms/1177729694}

\bibitem[{{Litwin-Kumar} \& Doiron(2012)}]{Litwin-Kumar2012_1498}
{Litwin-Kumar}, A., \& Doiron, B. 2012, Nature neuroscience, 15, 1498,
  \dodoi{10.1038/nn.3220}

\bibitem[{Mar{\v c}enko \& Pastur(1967)}]{Marcenko1967_457}
Mar{\v c}enko, V.~A., \& Pastur, L.~A. 1967, Mathematics of the USSR-Sbornik,
  1, 457, \dodoi{10.1070/SM1967v001n04ABEH001994}

\bibitem[{Mochizuki {et~al.}(2016)Mochizuki, Onaga, Shimazaki, Shimokawa,
  Tsubo, Kimura, Saiki, Sakai, Isomura, Fujisawa, Shibata, Hirai, Furuta,
  Kaneko, Takahashi, Nakazono, Ishino, Sakurai, Kitsukawa, Lee, Lee, Jung,
  Babul, Maldonado, Takahashi, {Arce-McShane}, Ross, Sessle, Hatsopoulos,
  Brochier, Riehle, Chorley, Grun, Nishijo, {Ichihara-Takeda}, Funahashi,
  Shima, Mushiake, Yamane, Tamura, Fujita, Inaba, Kawano, Kurkin, Fukushima,
  Kurata, Taira, Tsutsui, Ogawa, Komatsu, Koida, Toyama, Richmond, \&
  Shinomoto}]{Mochizuki2016_5736}
Mochizuki, Y., Onaga, T., Shimazaki, H., {et~al.} 2016, Journal of
  Neuroscience, 36, 5736, \dodoi{10.1523/JNEUROSCI.0230-16.2016}

\bibitem[{Mongillo {et~al.}(2018)Mongillo, Rumpel, \&
  Loewenstein}]{Mongillo2018_1463}
Mongillo, G., Rumpel, S., \& Loewenstein, Y. 2018, Nature Neuroscience, 21,
  1463, \dodoi{10.1038/s41593-018-0226-x}

\bibitem[{Muir \& {Mrsic-Flogel}(2015)}]{Muir2015_042808}
Muir, D.~R., \& {Mrsic-Flogel}, T. 2015, Physical Review E, 91, 042808,
  \dodoi{10.1103/PhysRevE.91.042808}

\bibitem[{Nakagawa \& Cuthill(2007)}]{Nakagawa2007_591a}
Nakagawa, S., \& Cuthill, I.~C. 2007, Biological Reviews, 82, 591,
  \dodoi{10.1111/j.1469-185X.2007.00027.x}

\bibitem[{Pernice {et~al.}(2011)Pernice, Staude, Cardanobile, \&
  Rotter}]{Pernice2011_e1002059}
Pernice, V., Staude, B., Cardanobile, S., \& Rotter, S. 2011, PLOS
  Computational Biology, 7, e1002059, \dodoi{10.1371/journal.pcbi.1002059}

\bibitem[{Rajan \& Abbott(2006)}]{Rajan2006_188104}
Rajan, K., \& Abbott, L.~F. 2006, Physical Review Letters, 97, 188104,
  \dodoi{10.1103/PhysRevLett.97.188104}

\bibitem[{Schlesinger(1979)}]{Schlesinger1979_103}
Schlesinger, S. 1979, Simulation, 32, 103, \dodoi{10.1177/003754977903200304}

\bibitem[{Schuessler {et~al.}(2020)Schuessler, Dubreuil, Mastrogiuseppe,
  Ostojic, \& Barak}]{Schuessler2020_013111}
Schuessler, F., Dubreuil, A., Mastrogiuseppe, F., Ostojic, S., \& Barak, O.
  2020, Physical Review Research, 2, 013111,
  \dodoi{10.1103/PhysRevResearch.2.013111}

\bibitem[{Sommers {et~al.}(1988)Sommers, Crisanti, Sompolinsky, \&
  Stein}]{Sommers1988_1895}
Sommers, H.~J., Crisanti, A., Sompolinsky, H., \& Stein, Y. 1988, Physical
  Review Letters, 60, 1895, \dodoi{10.1103/PhysRevLett.60.1895}

\bibitem[{Sompolinsky {et~al.}(1988)Sompolinsky, Crisanti, \&
  Sommers}]{Sompolinsky1988_259}
Sompolinsky, H., Crisanti, A., \& Sommers, H.~J. 1988, Physical Review Letters,
  61, 259, \dodoi{10.1103/PhysRevLett.61.259}

\bibitem[{Staude {et~al.}(2010)Staude, Rotter, \& Gr{\"u}n}]{Staude2010_327}
Staude, B., Rotter, S., \& Gr{\"u}n, S. 2010, Journal of Computational
  Neuroscience, 29, 327, \dodoi{10.1007/s10827-009-0195-x}

\bibitem[{Stella {et~al.}(2021)Stella, Bouss, Palm, \&
  Gr{\"u}n}]{Stella2021_2021.08.24.457480}
Stella, A., Bouss, P., Palm, G., \& Gr{\"u}n, S. 2021, Comparing Surrogates to
  Evaluate Precisely Timed Higher-Order Spike Correlations,
  \dodoi{10.1101/2021.08.24.457480}

\bibitem[{{Student}(1908)}]{Student1908_1}
{Student}. 1908, Biometrika, 6, 1, \dodoi{10.2307/2331554}

\bibitem[{Szucs \& Ioannidis(2017)}]{Szucs2017_}
Szucs, D., \& Ioannidis, J. P.~A. 2017, Frontiers in Human Neuroscience, 11,
  \dodoi{10.3389/fnhum.2017.00390}

\bibitem[{Trensch {et~al.}(2018)Trensch, Gutzen, Blundell, Denker, \&
  Morrison}]{Trensch2018_81}
Trensch, G., Gutzen, R., Blundell, I., Denker, M., \& Morrison, A. 2018,
  Frontiers in Neuroinformatics, 12, 81, \dodoi{10.3389/fninf.2018.00081}

\bibitem[{{van Albada} {et~al.}(2018){van Albada}, Rowley, Senk, Hopkins,
  Schmidt, Stokes, Lester, Diesmann, \& Furber}]{vanAlbada2018_291}
{van Albada}, S.~J., Rowley, A.~G., Senk, J., {et~al.} 2018, Frontiers in
  neuroscience, 12, 291, \dodoi{10.3389/fnins.2018.00291}

\bibitem[{Venkatesh {et~al.}(2019)Venkatesh, Jaja, \&
  Pessoa}]{Venkatesh2019_116398}
Venkatesh, M., Jaja, J., \& Pessoa, L. 2019, NeuroImage, 116398,
  \dodoi{10.1016/j.neuroimage.2019.116398}

\bibitem[{Vlachos {et~al.}(2012)Vlachos, Aertsen, \&
  Kumar}]{Vlachos2012_e1002311}
Vlachos, I., Aertsen, A., \& Kumar, A. 2012, PLOS Computational Biology, 8,
  e1002311, \dodoi{10.1371/journal.pcbi.1002311}

\bibitem[{Wainrib \& Touboul(2013)}]{Wainrib2013_118101}
Wainrib, G., \& Touboul, J. 2013, Physical Review Letters, 110, 118101,
  \dodoi{10.1103/PhysRevLett.110.118101}

\bibitem[{Zhou {et~al.}(2009)Zhou, Jin, \& Zhao}]{Zhou2009_2931}
Zhou, Q., Jin, T., \& Zhao, H. 2009, Neural Computation, 21, 2931,
  \dodoi{10.1162/neco.2009.12-07-671}

\end{thebibliography}

\bibliographystyle{aasjournal}

\section*{Supplements}
\subsection*{Data \& code availability}
\begin{itemize}
    \item All code for the simulations, analysis, and visualizations are stored in the gin repository. \url{https://gin.g-node.org/INM-6/eigenangles}
    \item All simulated and processed data is also stored alongside the code in the same gin repository. \url{https://gin.g-node.org/INM-6/eigenangles}
    \item The eigenangle test to compare correlation matrices will be provided as part of the v0.2 release of the Python validation test library NetworkUnit  \url{https://github.com/INM-6/NetworkUnit} (RRID:SCR\_016543).
\end{itemize}

\subsection*{Interactive notebook of the test construction}
An interactive jupyter notebook illustrates the construction of the eigenangle similarity measure and its statistical evaluation with a combination of text, code snippets, and interactive figures: \url{https://gin.g-node.org/INM-6/eigenangles/eigenangle_basics.ipynb}\footnote{executable in the browser via \url{https://mybinder.org/v2/git/https\%3A\%2F\%2Fgin.g-node.org\%2FINM-6\%2Feigenangles/HEAD?labpath=eigenangle_basics.ipynb}}

\subsection*{Supplementary figures} 
\begin{suppfigure}[!h]
	\centering
	\includegraphics[width=.85\textwidth]{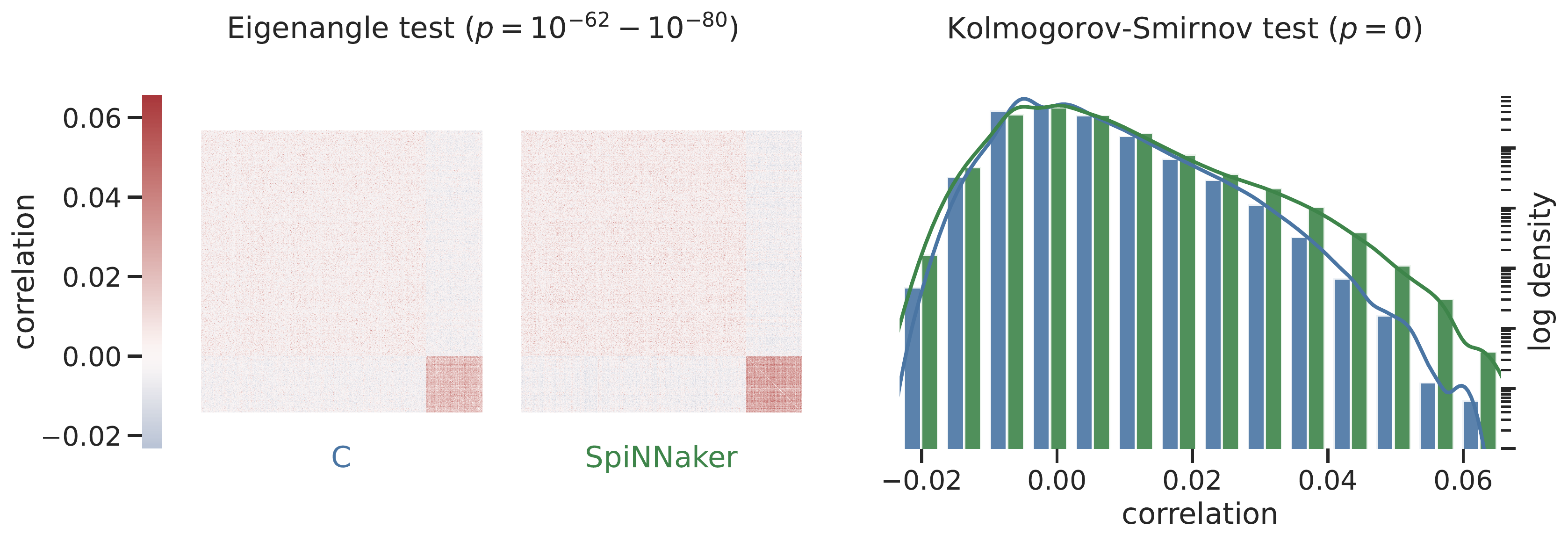}
	  \caption{\textbf{Comparing the model activity from two simulators.} The Izhikevich polychronization model \citep{Izhikevich2006_245} was simulated \citep{Trensch2018_81, Gutzen2018_90} with identical parameters and initial conditions on both a custom simulator written in C (blue) and the neuromorphic system SpiNNaker (green). Here, the correlation matrices (\textit{left}) and distributions of correlation coefficient (\textit{right}) from a $60$~s recording are shown. Comparing $5$ recordings, both the eigenangle test and the KS test clearly reject their respective null hypothesis, indicating a similarity of the correlation structure by the eigenangle test while the KS test indicates dissimilarity of the distributions of correlation coefficients.}\label{suppfig:simulator_comparison}
\end{suppfigure}

\begin{table}
\caption{Network configuration}\label{tab:network_configuration}
\begin{tabular}{lll}
\toprule
parameter & value & description  \\ 
\toprule
$N$ & $1000$ & number of neurons\\
$f$ & $0.8$ & fraction of exc. neurons\\
$J_{ex}$ & $0.1$ & mean exc. strength [mV]\\
$J_{in}$ & $-f J_{ex}/ (1-f)$& mean inh. strength [mV]\\
weight distribution & lognormal & distribution to sample synaptic strengths \\
$\sigma_{ex}$ & $0.12$ & standard deviation of exc. weight distribution [mV]\\
$\sigma_{in}$ & $0.1$ & standard deviation of inh. weight distribution [mV]\\
$T$ & $60000$ & simulation time [ms]\\
$T_0$ & $1000$ & disregarded swinging-in time (additional to $T$) [ms]\\
$\epsilon$ & $0.1$ & connection probability\\ 
$\eta$ & $0.9$ & external rate relative to threshold rate\\
delay & uniform(min=$0.5$, max=$3.0$) & distribution to sample synaptic delays [ms] \\
$dt$ & $0.1$ & time resolution [ms]\\
connection rule & pairwise Bernoulli & rule for connection neurons\\
synapse model & static &  type of synaptic connection\\
neuron model & current-based, delta, leaky iaf & type of neuron model\\
$\tau_m$ & $20.0$ & time constant of membrane potential [ms] \\
$\theta$ & $20.0$ & membrane threshold potential [mV]\\
$C_{m}$ & $1.0$ & membrane capacitance [pF]\\
$t_{ref}$ & $2.0$ & duration of refractory period [ms]\\
$E_L$ & $0.0$ & resting membrane potential [mV]\\
$V_{reset}$ & $0.0$ & reset potential of the membrane [mV]\\
$V_m$ & $0.0$ & membrane potential [mV]\\
stimulus type & independent Poisson & driving stimulus to all neurons\\
stimulus rate & $1000 \cdot \eta \cdot \theta / (J_{ex}\tau_m)$& average rate of the Poisson generator [Hz]\\
\toprule
\end{tabular}
\end{table}

\begin{suppfigure}[!h]
	\centering
	\includegraphics[width=.85\textwidth]{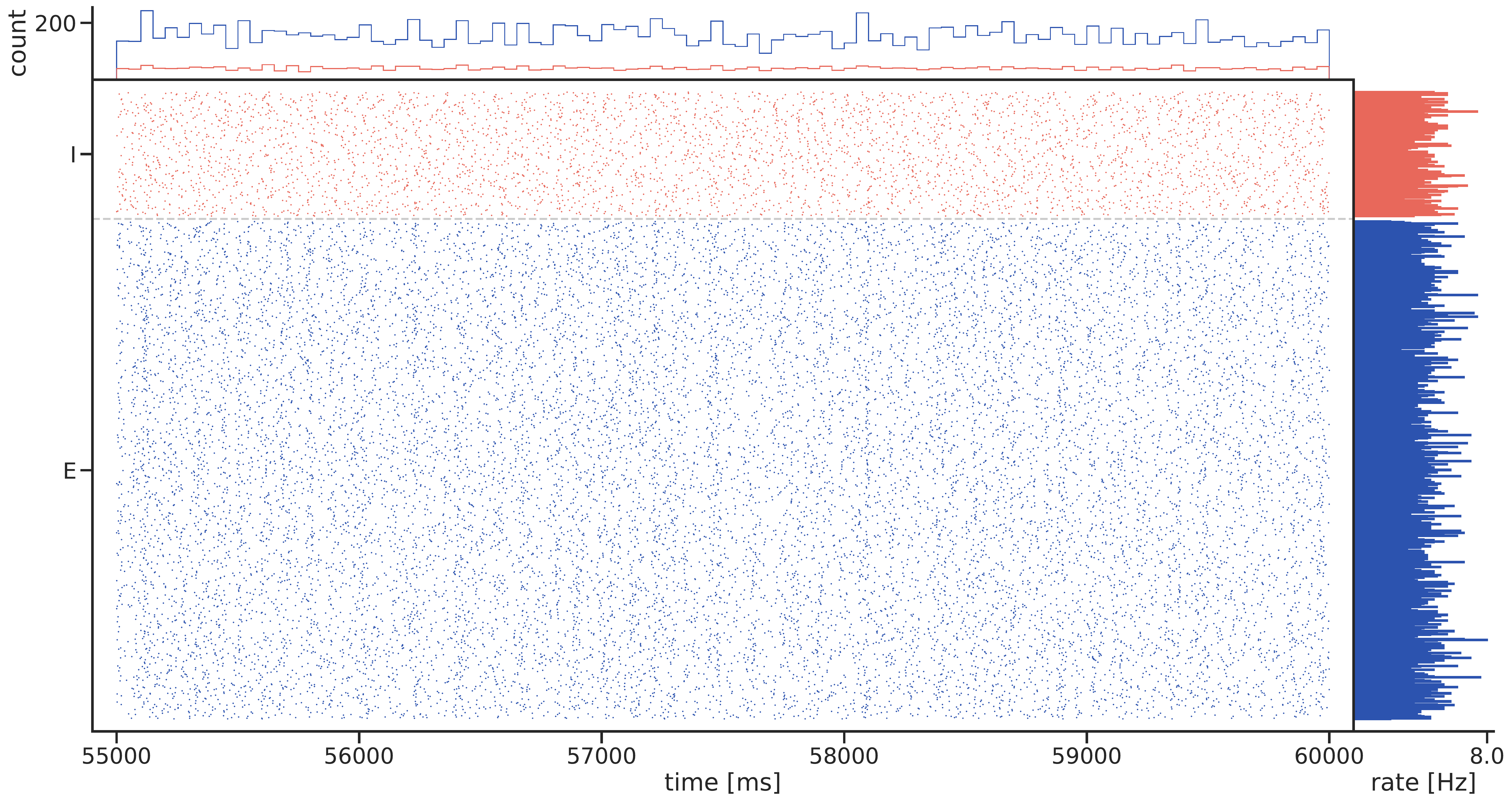}
	\caption{\textbf{Spiking activity of network model.} The raster plot shows the simulated spiking activity of all $1000$ neurons, $800$ excitatory (E, blue) and $200$ inhibitory (I, red), of the network model (see \prettyref{sec:asymmetric_extension} and \prettyref{tab:network_configuration}) in a $5$~s window. Top: population histogram for each population ($50$~ms bins). Right: mean firing rate for each neuron.}
	\label{suppfig:rasterplot}
\end{suppfigure}

\begin{suppfigure}[!h]
	\centering
		\includegraphics[width=.95\textwidth]{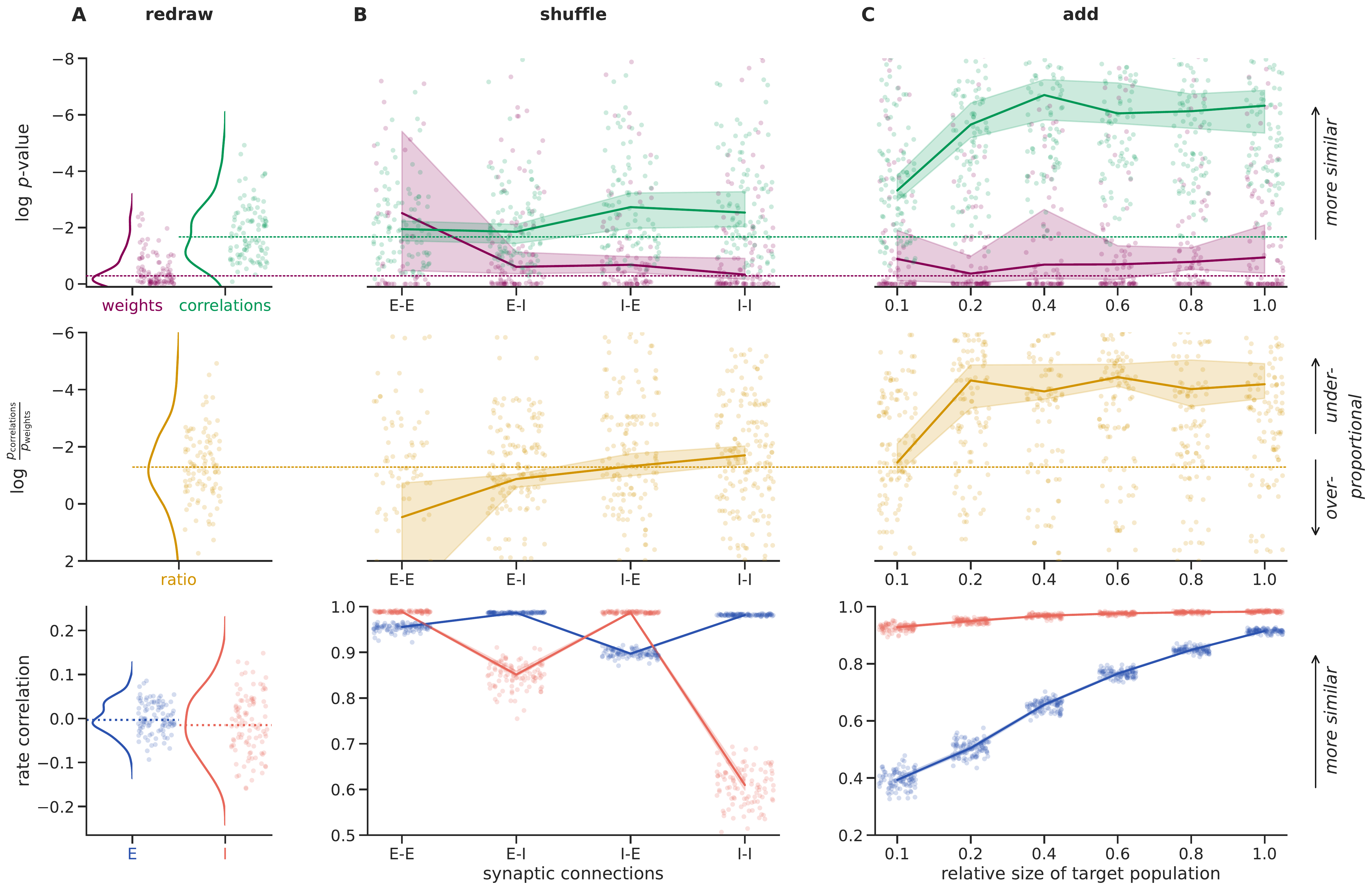}
	  \caption{\textbf{Effects of network rewiring with same numbers of changes synapses.} Same as \prettyref{fig:rewiring_comparison}, but for the \textit{shuffle} protocol the same number of synapses ($3500$) are rewired for E-E, E-I, I-E, and I-I; and for the \textit{add} protocol the same number of synapses ($12800$, corresponds to $20\%$ of E-E connections) are added for each size of the target population.
	  }
	  \label{suppfig:rewire_comparisons_synapse_corrected}
\end{suppfigure}



\end{document}